\documentclass[sn-basic,iicol,dvipsnames,table]{sn-jnl}% Basic Springer Nature Reference Style/Chemistry Reference Style
\usepackage[version=3]{mhchem}
\usepackage{balance}
\usepackage{mathptmx}
\usepackage{graphicx} 
\usepackage{subcaption}
\usepackage[format=plain,justification=justified,singlelinecheck=false,font={stretch=1.125,small,sf},labelfont=bf,labelsep=space]{caption}
\usepackage{float}
\usepackage{fancyhdr}
\usepackage{fnpos}
\usepackage[english]{babel}
\usepackage{array}
\usepackage{droidsans}
\usepackage{charter}
\usepackage[T1]{fontenc}
\usepackage{xcolor}
\usepackage{setspace}
\usepackage{hyperref}
\usepackage{bbding}
\usepackage{adjustbox}
\usepackage{textcomp}
\usepackage{soul}

\usepackage{siunitx}
\usepackage{amsmath,amssymb}

\usepackage{epstopdf}%This line makes .eps figures into .pdf - please comment out if not required.

\usepackage{microtype}

\jyear{2021}%

\theoremstyle{thmstyleone}%
\theoremstyle{thmstyletwo}%

\theoremstyle{thmstylethree}%

\begin{document}

\title{Rheological design of thickened alcohol-based hand rubs}

%%=============================================================%%
%% Prefix	-> \pfx{Dr}
%% GivenName	-> \fnm{Joergen W.}
%% Particle	-> \spfx{van der} -> surname prefix
%% FamilyName	-> \sur{Ploeg}
%% Suffix	-> \sfx{IV}
%% NatureName	-> \tanm{Poet Laureate} -> Title after name
%% Degrees	-> \dgr{MSc, PhD}
%% \author*[1,2]{\pfx{Dr} \fnm{Joergen W.} \spfx{van der} \sur{Ploeg} \sfx{IV} \tanm{Poet Laureate} 
%%                 \dgr{MSc, PhD}}\email{iauthor@gmail.com}
%%=============================================================%%

\author*[1]{\fnm{Andreia F} \sur{Silva}}\email{andreia.silva@ed.ac.uk}

\author[1]{\fnm{Tiffany A} \sur{Wood}}

\author[1]{\fnm{Daniel J M} \sur{Hodgson}}

\author[1]{\fnm{John R} \sur{Royer}}

\author[1]{\fnm{Job H J} \sur{Thijssen}}

\author[1]{\fnm{Alex} \sur{Lips}}

\author*[1]{\fnm{Wilson C K} \sur{Poon}}\email{w.poon@ed.ac.uk}

\affil[1]{Edinburgh Complex Fluids Partnership (ECFP), SUPA and School of Physics \& Astronomy, The University of Edinburgh, Peter Guthrie Tait Road, Edinburgh EH9 3FD, United Kingdom}

%%==================================%%
%% sample for unstructured abstract %%
%%==================================%%

\abstract{The handleability and sensory perception of hand sanitisers by consumers affect the hygiene outcome. Spillage may result in under-dosing and poor sensory properties can lead to under-utilisation. We first propose four principles (low run off, spreadability, smoothness and non-stickiness) for designing the rheology of thickened alcohol-based hand rubs with acceptable handleability and hand feel. We then evaluate a commercial hand gel and a variety of simplified formulations thickened with microgels (Carbopol 974P, Carbopol Ultrez 20 and Sepimax Zen), or linear polymers (Jaguar HP 120 COS), and evaluate them against these design criteria. All four additives provide acceptable spreadability by shear thinning to $\eta \approx \SI{e-1}{\pascal\second}$ at $\dot\gamma \sim \SI{e3}{\per\second}$. Either the finite yield stress conferred by the microgels ($\sigma_y \gtrsim \SI{10}{\pascal}$) or the increase in low-shear viscosity provided by the linear polymer ($\eta \gtrsim \SI{1}{\pascal\second}$ at $\dot\gamma \lesssim \SI{0.1}{\per\second}$) give rise to acceptably low run-off. However, the formulation using the linear polymer shows a filament breakage time of $\tau_{\rm b} \approx \SI{1}{\second}$ in capillary rheology, which may result in stickiness and therefore a less than optimal hand feel.}

%%================================%%
%% Sample for structured abstract %%
%%================================%%

\keywords{Hand sanitisers, Microgels, Polymers, Rheology}

%%\pacs[JEL Classification]{D8, H51}

%%\pacs[MSC Classification]{35A01, 65L10, 65L12, 65L20, 65L70}

\maketitle

\section{Introduction}

The COVID-19 pandemic has brought hand sanitising to the forefront of public attention. The sudden rise in demand in early 2020 for hydroalcoholic hand gels, commonly known as alcohol hand gels, soon outstripped supply. Users turned to liquid formulations recommended by the \cite{WHO} (WHO) consisting of alcohol (ethanol or propanol at 80\% and 75\% by volume respectively) and water mixed with small amounts of hydrogen peroxide and glycerol. In parallel, a range of manufacturers scrambled to bring to market new thickened alcohol-based hand rubs (ABHRs), focussing unprecedented attention on how to formulate such products. The high demand for thickened ABHRs will likely persist post-pandemic into the `new normal'. 

Early on in the pandemic, \cite{Berardi2020} published a survey of ABHRs. They measured the viscosity of 17 products on the Italian market as of April 2020 as a function of shear rate, $\eta(\dot\gamma)$. The majority show a finite yield stress (see their supplementary Fig.~SF1 inset). After reviewing thickeners, dissolution and regulation, the authors gave a practical guide for ingredient selection. 

\cite{Berardi2020} only briefly discuss how easy the products are to handle and their hand feel. A review by \cite{Greenaway2018} focuses on these issues. They report that the `poor handleability' of WHO-type liquid hand sanitisers lead to spillage and under-dosing, and that users' `acceptance of the sensory properties of ABHRs during and after application' affects the hygiene outcome. 

More recently, \cite{Villa2021} have reviewed the most commonly-used polymeric thickeners. They compiled a list of alcohol-based hydrogel formulations based on suppliers' information. The authors point out that while these thickeners in aqueous media have been studied, their rheology in hydroalcoholic solvents have not yet been investigated in any depth. 

The rheology of an ABHR is a key determinant of its handleability and hand feel, although many other factors (e.g.~the time-dependent water activity during drying) control the latter. Yet, the rheology of these (or, indeed, any other topical application) products is usually arrived at through trial and error. In this work, we set out design principles for thickened ABHRs to deliver acceptable handleability and hand feel based on what is known about the soft-matter science of the material systems. We then evaluate a commercial ABHR against these criteria, and explore the use of four thickeners in experimental minimal formulations against our criteria. Other areas of `pandemic soft matter science' have been reviewed elsewhere \citep{Poon2020}.

\section{Designing thickened hand sanitiser rheology}\label{sec:design}

We first propose, based on existing fundamental understanding and a few empirical observations, a number of design principles for the rheology of thickened ABHRs.

\subsection{Preventing run off}

A key reason for thickening liquid-like hand sanitisers is to improve handleability. Consider a volume $V = \SI{2}{\milli\litre}$ of WHO hand sanitising liquid deposited onto a palm. This amount is needed to provide enough hand coverage to reduce microbial contamination by a factor of $10^2$, which is the US FDA efficacy criterion for such products \citep{Kampf2013}. Experience suggests, and a simple calculation (see Appendix~1) confirms, that a \SI{2}{\milli\litre} dose of a \SI{2}{\milli\pascal\second} Newtonian liquid (see Sec.~\ref{sec:WHOrheol}) takes $\lesssim \SI{e-1}{\second}$ to run off a palm inclined at $\approx \SI{20}{\degree}$, so that the WHO formulations indeed have `poor handleability' \citep{Greenaway2018}.

Most manufacturers solve this by using a polymeric thickener to turn the solution into a gel with finite yield stress, $\sigma_y$. A sessile drop of height $h \lesssim \SI{1}{\centi\meter}$ and density $\rho \lesssim \SI{e3}{\kilo\gram\per\cubic\meter}$ on a palm inclined at angle $\alpha$ experiences a shear stress $\sigma \sim \frac{\rho g h}{2} \sin\alpha$ (with $g$ the gravitational acceleration), or $\sigma \gtrsim \SI{10}{\pascal}$ for $\theta = \SI{20}{\degree}$. A gelled ABHR should therefore have $\sigma_y \gtrsim \SI{10}{\pascal}$.

However, the processing and bottling of yield-stress fluids is in general more difficult than Newtonian liquids. Moreover, alcohol-water mixtures evaporate rapidly \citep{Prash2020}, and $\sigma_y$ increases with thickener concentration \citep{Kim2003}. Thus, dried residual material can clog dispenser nozzles. (In a slightly different context, such clogging is well-known for nozzles extruding hydrogels for 3D printing \citep{Li2018}.) So, is a finite $\sigma_y$ necessary for handleability?

The run off speed of films \citep{Batchelor:1967ay} and droplets \citep{Kim2002} scales as the inverse fluid viscosity. A factor of $\sim 10^3$ increase in the viscosity in an ABHR compared to the WHO formulation should therefore  reduce the run-off time of a \SI{2}{\milli\litre} dose to $\gtrsim \SI{1}{\minute}$. Thus, $\eta \gtrsim \SI{1}{\pascal\second}$ at $\dot\gamma \lesssim \SI{0.1}{\per\second}$ should suffice to confer reasonable handleability to a thickened hand sanitiser without yield stress. 

\subsection{Hand feel}

Hand feel depends on both rheology and physicochemical determinants such as water activity, the latter controlling  how `moisturising' a product feels and how rapidly it dries. We focus on the rheological aspects. 

Most topical formulations shear thin, which users perceive as `spreadability' \citep{Kwak2015}. The degree of shear thinning needed is dictated by the desired viscosity at $\dot\gamma \gtrsim \SI{e3}{\per\second}$. Users typically rub topical products to a thickness of $\sim \SI{20}{\micro\meter}$. This can be rationalised by experiments rubbing a glass sphere on human skin \citep{Adams2007}, which find that a film of $\approx $\SI{10}{\micro\meter} is needed to separate (rough) skin from (smooth) glass. Thus, a film of $\approx \SI{20}{\micro\meter}$ completely separates two skin surfaces, which is apparently what users desire. So at a final rubbing speed of 1-\SI{10}{\centi\meter\per\second}, $\SI{e3}{\per\second} \lesssim \dot\gamma \lesssim \SI{e4}{\per\second}$. Slightly extrapolating literature data \citep{Kwak2015}, we find viscosities at such $\dot\gamma$ of \SI{e-1}{\pascal\second} (lotions) to \SI{1}{\pascal\second} (creams). So, a thickened ABHR should thin to $\eta \gtrsim \SI{e-1}{\pascal\second}$ at $\dot\gamma \gtrsim \SI{e3}{\per\second}$.

Adding polymer thickeners potentially confers elasticity, and therefore normal stress differences, on formulations. The role of first normal stress difference, $N_1$, at high rates of deformation is debated. Some suggest that it can confer the feeling of `smoothness' or `moistness' in topical products in the final stages of rubbing \citep{Tamura2013}. Another potential design criterion is therefore the development of measurable $N_1$ at $\dot\gamma \sim \SI{e3}{\per\second}$.

However, elasticity also confers `spinnability'. The basic experiment here is that of `finger extensional rheology': stretching a formulation between the thumb and index finger. An elastic formulation behaves like saliva, viz., forms long threads that take a perceptible time to break up. It is claimed that such `spinnability' is experienced as `stickiness' in the mouth or on skin \citep{Tamura2013,Dinic2019}. Interestingly, data implicitly demonstrating this correlation exist \citep{Wolf2016}. We make this explicit in Appendix~2, and use this published data to suggest that a filament breakage time of $\tau_{\rm b} \lesssim \SI{1}{\second}$ can be a rational criterion for `non-stickiness'.

\subsection{Summary}

Our design principles for thickened ABHRs are:
\begin{itemize}
	\item Low runoff: $\sigma_y \gtrsim \SI{10}{\pascal}$ or $\eta \gtrsim \SI{1}{\pascal\second}$ at $\dot\gamma \lesssim \SI{0.1}{\per\second}$
	\item Spreadability at $\SI{20}{\micro\meter}$: shear thins to $\eta \approx \SI{e-1}{\pascal\second}$ at $\dot\gamma \sim \SI{e3}{\per\second}$
	\item Smoothness: significant $N_1$ at $\dot\gamma \sim \SI{e3}{\per\second}$ to prevent direct skin-skin contact
	\item Not sticky: filament breakage time of $\tau_{\rm b} \lesssim \SI{1}{\second}$
\end{itemize}

\section{Materials and methods}

In this work, we measure the rheology of a commercial product and formulations thickened with four different polymers and discuss their performance vis-\`a-vis these principles.

\subsection{Materials}

Purell `Advanced Hygenic Hand Rub', which contains 70\%  (v/v) ethanol thickened by a hydrophobised carbomer, was used as purchased. We compared its rheology against a WHO formulation on its own and thickened by various commercial polymers. These polymers' precise compositions are not publicly available, but some chemical information is available from their International Nomenclature of Cosmetic Ingredients (INCI) name. 

Many ABHRs are thickened by carbomers \citep{Berardi2020,Brady2017}, which are microgel particles (first produced by B F Goodrich, now Lubrizol, as Carbopol$^{\mbox{\textregistered}}$) of crossed-linked networks of copolymers of acrylic acid (56-68\% w/w) and alkyl-methacrylate. Microgels thicken solutions by swelling to their jamming point and beyond, developing a finite yield stress \citep{Angelini2018}. The commercial product we study is thickened by a carbomer with INCI name `Acrylates/C10-30 alkyl acrylate crosspolymer'. A common product with this INCI name is Lubrizol's Carbopol$^{\mbox{\textregistered}}$ Ultrez~20, a hydrophobically-modified carbomer \citep{Ultrez}. Invented to withstand high electrolyte concentration, these carbomers also disperse easily in alcohol. We study a minimal formulation of alcohol, water,  Ultrez~20 and small amounts of hydrogen peroxide and glycerol.

For comparison, we study two other hydrophobically-modified thickening agents. Sepimax Zen$^{\rm TM}$ (from Seppic; INCI name `Polyacrylate Crosspolymer-6') \citep{Sepimax} is a polymerised mixture of acryolyldimethyltaurate (sulphonate-bearing) monomers with a variety of polyacrylic acid-bearing monomers of which some are significantly hydrophobised. Preliminary mass spectrometry gave a molecular weight of $\sim \SI{e5}{\dalton}$ \citep{Crosby2}.  Jaguar\textsuperscript{\textregistered} HP 120 COS (Solvay; INCI name `hydroxypropyl guar gum') is a hydrophobised natural polysaccharide \citep{Lapasin1995,Cheng2002}.

Finally, for contrast, we compare these minimal formulations thickened with hydrophobically-modified agents with one that is thickened with a non-hydrophobically-modified agent, Carbopol$^{\mbox{\textregistered}}$~974P, a  low-residual-solvent carbomer (from Lubrizol, INCI name `Carbomer'; molecular weight  between $10^5$ and $\SI{e9}{\dalton}$) \citep{Schatz2015} widely used for thickening water-based formulations. 

Polymers were used as received. Ethanol (99.8\%), glycerol (98\%), hydrogen peroxide (30\%), triethanolamine ($\geq$99.0\%), triethylamine ($\geq$99\%) and citric acid ($\geq$99.5\%) from Sigma Aldrich were also used as received.

\subsection{Sample preparation}

\subsubsection{WHO formula} 

We added 0.125\% (v/v) \ce{H2O2} and 1.45\% (v/v) glycerol to 80\% ethanol and topped up the mixture with distilled water to 100\% (v/v). 

\subsubsection{Carbomers}

Carbomers dissolve when `neutralised' by alkali to generate ionised carboxylic groups \citep{Katdare2006}; the resulting strong electrostatic swelling produces a jammed aqueous gel \citep{Frisken2006}. The effect of alcohol is not understood, but likely involves differential alcohol/water adsorption on the polymer \citep{Kremer2014} and subtle counterion effects \citep{Sappidi2016,Gupta2017,Nishiyama2000a,Nishiyama2000b}.

Different concentrations (0.25, 0.35 and 0.5 \% (w/v)) of Ultrez 20 or Carbopol 974P were added directly to the solvent (WHO formula), vortex-mixed for $\approx \SI{15}{\second}$ to avoid clumping, and roller-mixed overnight before pH-adjusted using triethylamine to between 7 and 8. The pH was monitored using a meter equipped with a \ce{KCl} glass electrode (SevenExcellence S975-K, Mettler Toledo)  \citep{Bates1963}. Triethanolamine failed as pH modifier  at $>~70\%$ ethanol. Finally, the samples were pre-mixed using a vortex mixer for $\approx \SI{15}{\second}$ and roller-mixed for another night to maximise dissolution. 

\subsubsection{Sepimax Zen gels}

Sepimax Zen, also based on polyacrylic acid, is `pre-neutralised'. Different concentrations (0.5, 1 and 1.5\% (w/v)) were dispersed by overnight roller mixing with solvent  (WHO formula) without pH adjustment to a final pH between 5 and 6.5.

\subsubsection{Jaguar HP 120 COS gels}
\label{Jaguar_prot}

Jaguar HP 120 COS polymer (0.5, 1 and 1.5\% (w/v)) was added to the WHO formula and magnetically-stirred for \SI{2}{\hour} at 170~rpm. The pH was then adjusted to $\approx 5$ with citric acid (50\% (w/v)) and the samples were stirred overnight at 170~rpm.

\subsubsection{Alternative protocol}

Dissolving the polymers in water before adding the other ingredients (ethanol, hydrogen peroxide and glycerol) made no significant difference to the measured rheology. So, we do not report findings using this alternative protocol.

\subsection{Spectrophotometry}

Transmittance ($400-\SI{700}{\nano\meter}$) was measured in \SI{1}{\centi\meter} pathlength cells using a Cary 300 spectrophotometer (Agilent). 

\subsection{Steady shear rheology}

Steady shear flow curves were measured in a DHR-2 rheometer (TA Instruments) at \SI{20}{\celsius} under controlled rate using a sand-blasted cone-and-plate  (40 mm diameter, \SI{1}{\degree}) geometry. A \SI{40}{\milli\meter}-diameter cross-hatched plate (\SI{1}{\milli\meter} hatching) gave similar results, with no signs of slip in either geometry, so we report only the cone plate results. When appropriate, we fit the Herschel-Bulkley (HB) model $\sigma = \sigma_y + k\dot\gamma^n$ to obtain the yield stress $\sigma_y$.

The first normal stress difference, $N_1$, was measured using the cone-plate geometry \citep{morrison2001}, waiting 2 to \SI{10}{\minute} after filling the gap before starting each experiment to minimise residual axial force due to loading. 

A solvent trap was used in all steady shear measurements to  minimise evaporation.

\subsection{Capillary rheology}

We used a CaBER${^{\text{TM}}}$ device (Haake) to quantify the sample's filaments breakup time. A  \SI{50}{\milli\second} step with Hencky strain $\varepsilon = \ln(h_{\rm f}/h_{\rm i}) = 1.36$ was imposed, where $h_{\rm i}$ and $h_{\rm f}$ are the initial and the final gap heights, and the subsequent filament diameter  monitored at 20-\SI{22}{\celsius}. The evolution of the thinning fluid filament was measured using either the CaBER laser or a high speed camera (FASTCAM SA6, model 75K-M1). The video images were afterwards analysed with Matlab (MathWorks, version R2018a). For the imposed Hencky strain $h_{\rm f} \approx 4 h_{\rm i}$, mimicking a `finger test' of taking a film of a few \si{\milli\meter} between thumb and index finger and suddenly increasing the distance to between 1 and \SI{2}{\centi\meter}. We measured within 30 s of loading, during which evaporation was negligible ($<1.5\%$).

\section{Results}

\subsection{Optical quality}

\begin{figure}[t]
	\centering
	\subfloat[\centering]{\includegraphics[width=0.3\textwidth]{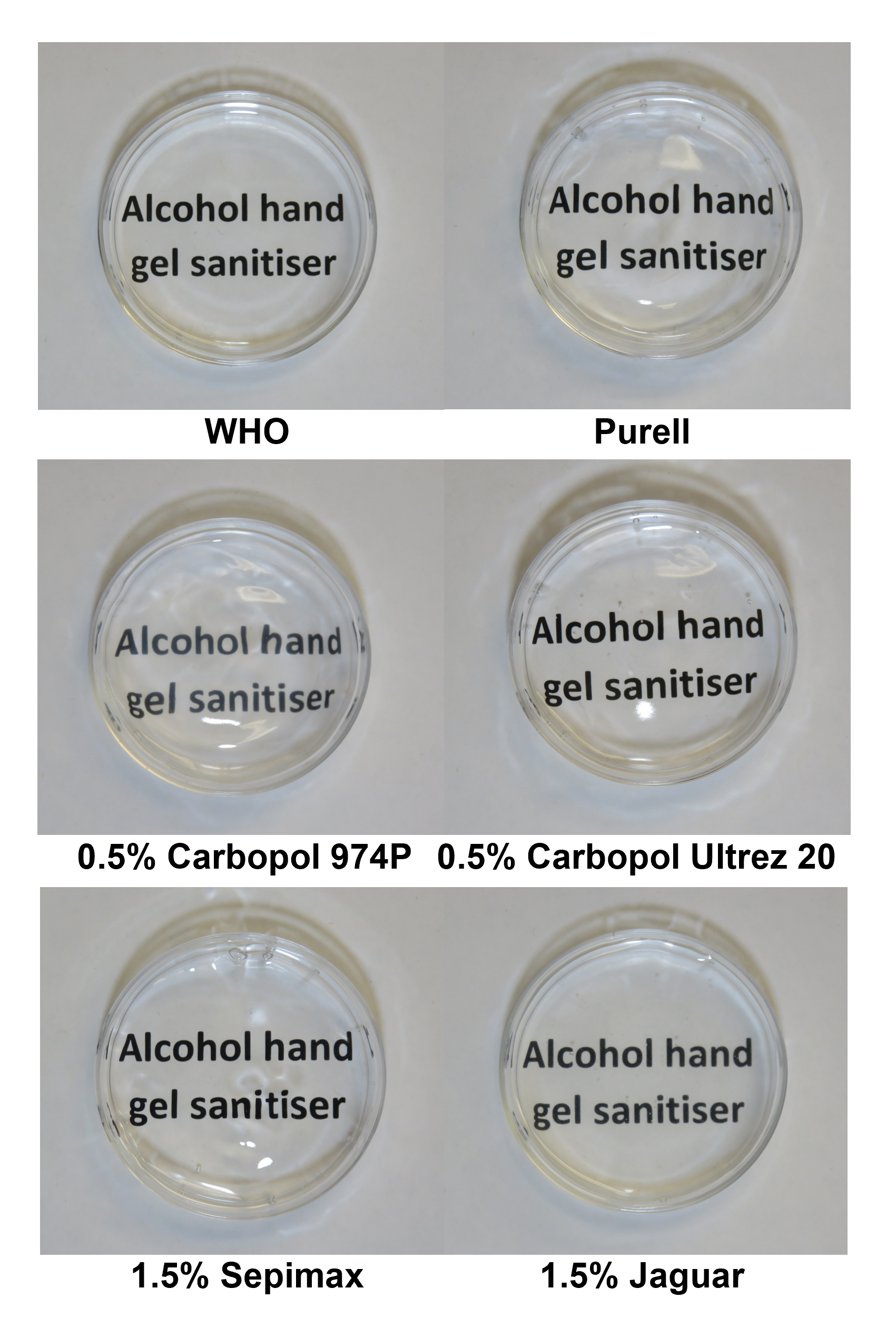}  \label{fig:gels_pic} }\\
	\subfloat[\centering]{\includegraphics[width=0.4\textwidth]{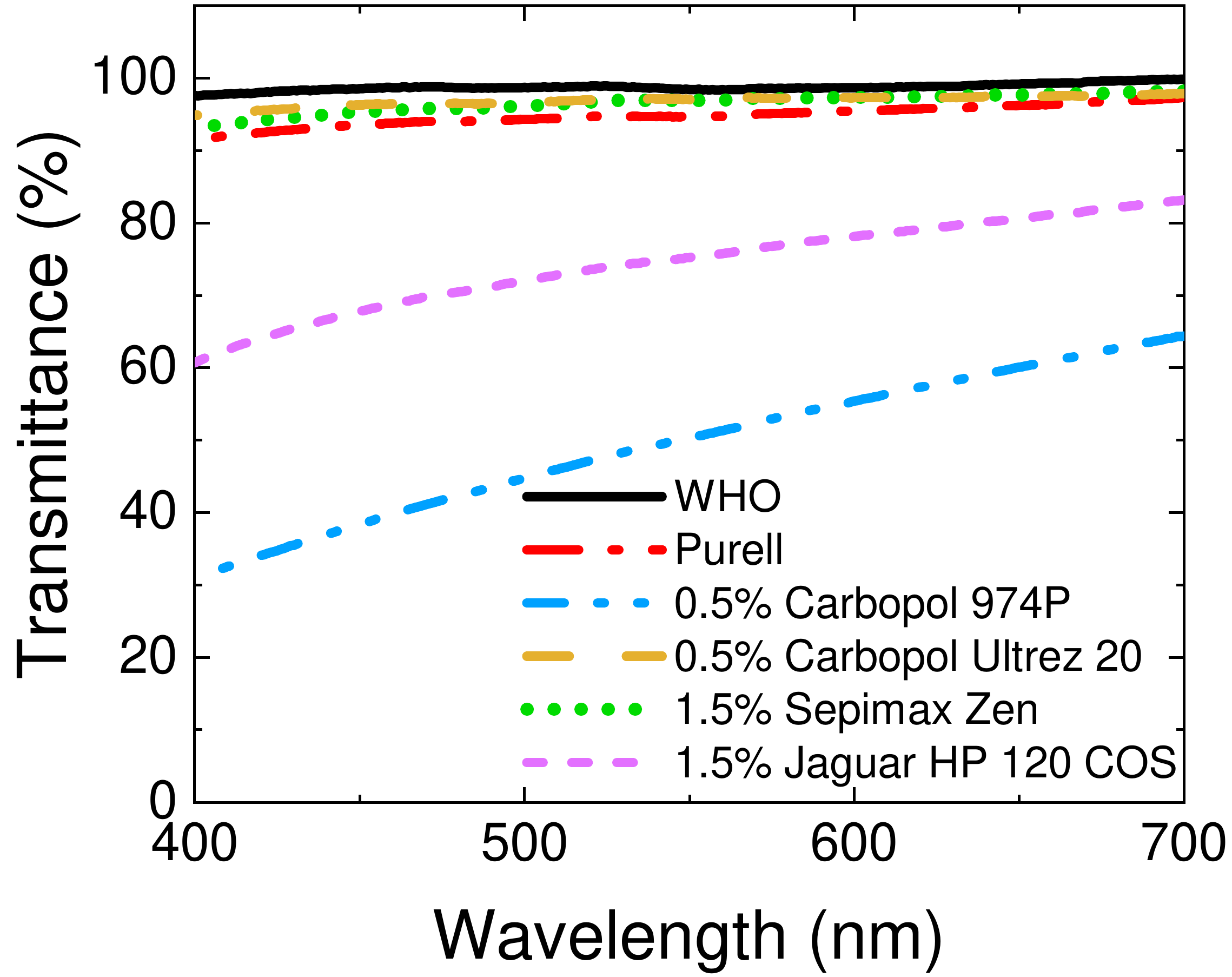} \label{fig:gels_transmittance}}
	\caption{(a) Visual assessment of transparency in \SI{4.5}{\centi\meter} Petri dishes filled to 1 cm atop printed text at polymer concentrations that give comparable rheology to the Purell gel. (b) Transmittance of the different ABHR formulations measured from 400 to $\SI{700}{\nano\meter}$. We show results at the highest concentration of each polymer;  lower concentration samples show similar or slightly higher transmittance.}
\end{figure}

We briefly assessed the optical quality of the most concentrated sample of each thickener, which all have comparable rheology to the commercial sample, by filling a  \SI{4.5}{\centi\meter} Petri dish to \SI{1}{\centi\meter} depth and inspecting atop lines of printed text, Fig.~\ref{fig:gels_pic}. All samples, the WHO formulation and the commercial product are transparent, although the samples thickened with Carbopol 974P and Jaguar HP 120 COS do appear a little less clear than the others. 

 Spectrophotometry, Fig.~\ref{fig:gels_transmittance}, confirms that gels containing Carbpol 974P and Jaguar HP 120 COS are more turbid than the other samples in the range  400-\SI{700}{\nano\meter}. This may indicate lower solubility of these two polymers, leaving undissolved aggregates to scatter light; but pursuing this further is beyond our scope \citep{Kremer2014}.

\subsection{Rheology}

\subsubsection{WHO liquid and commercial gel} \label{sec:WHOrheol}

\begin{figure}[]
	\centering
	\subfloat[\centering]{\includegraphics[width=0.255\textwidth]{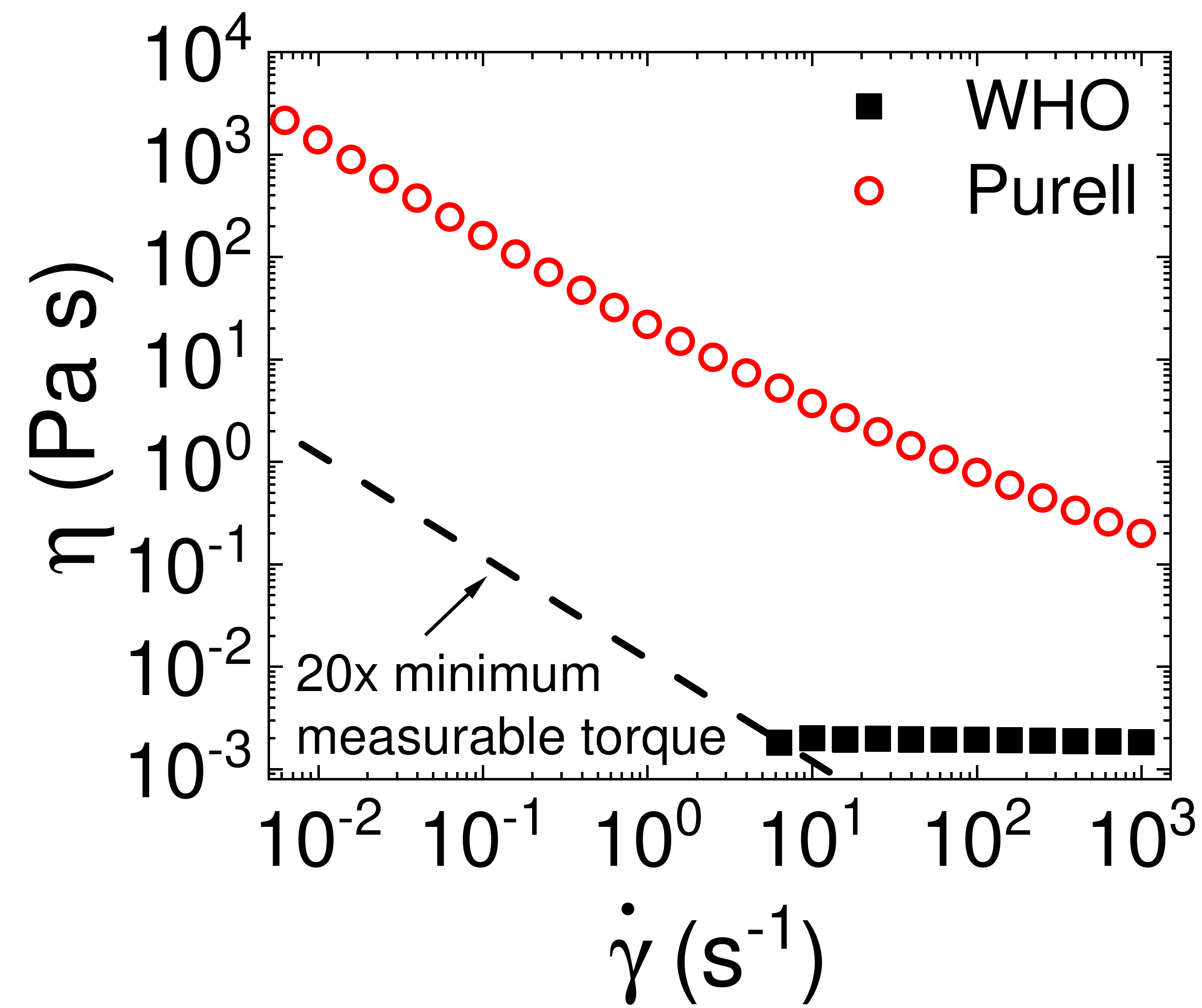}\label{fig:WHO_Purell_visc}}
	\subfloat[\centering]{\includegraphics[width=0.255\textwidth]{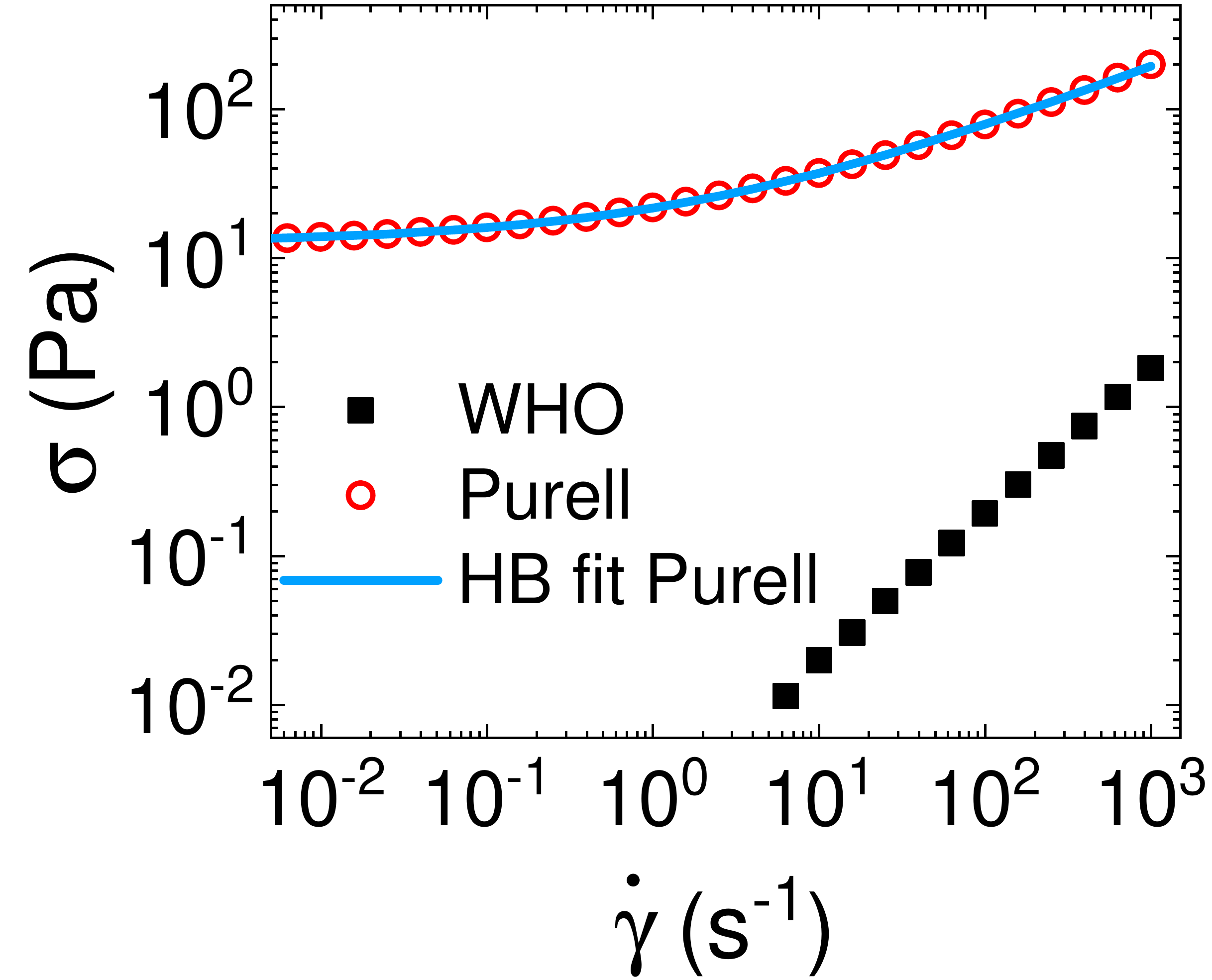}\label{fig:WHO_Purell_stress}}\\
	\subfloat[\centering]{\includegraphics[width=0.255\textwidth]{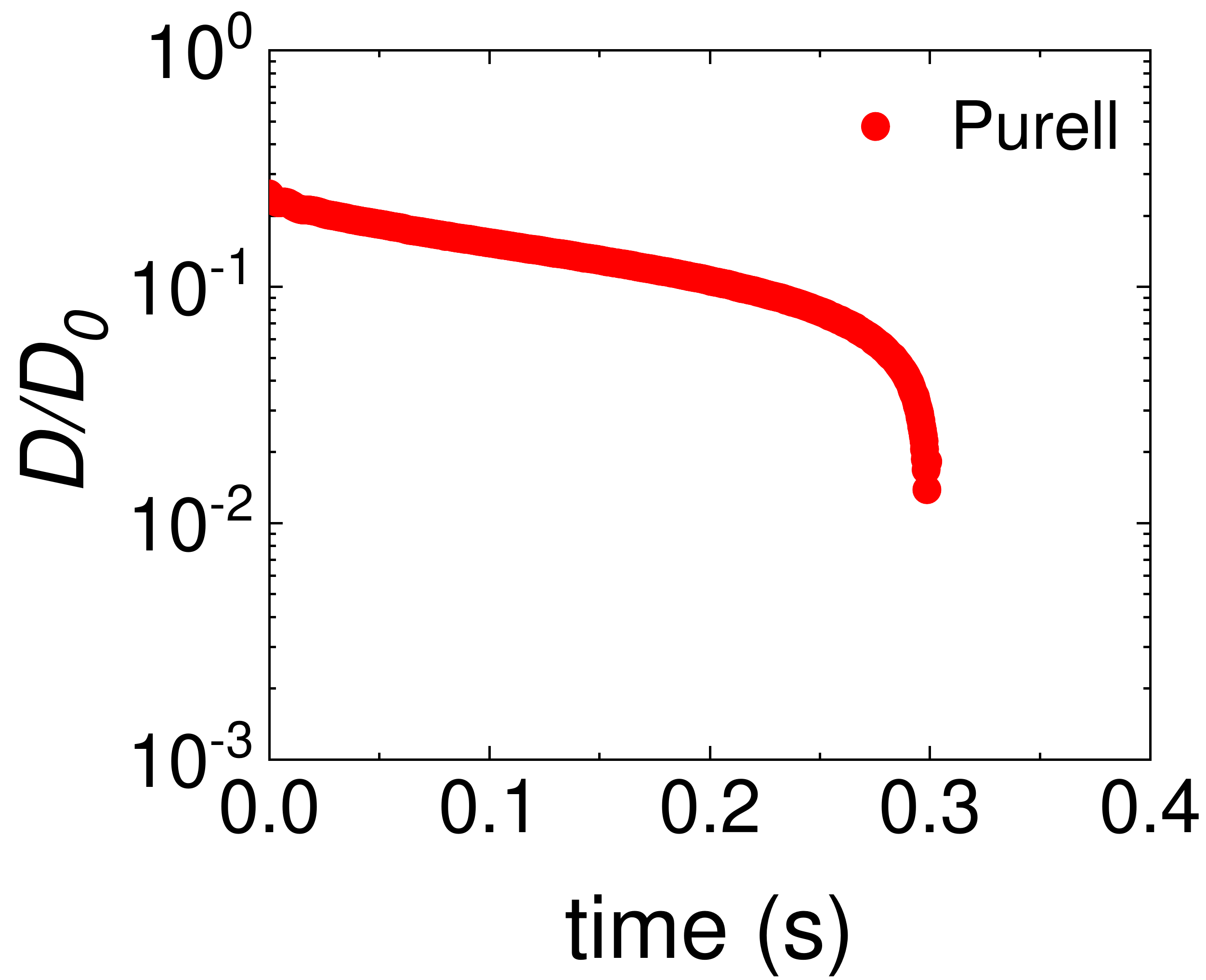}\label{fig:Purell_filament}}
	\caption{(a) Viscosity and (b) stress vs shear rate of the WHO 80\% formulation and commercial Purell hand gel. The dashed line in (a) is the lowest measurable shear viscosity corresponding to 20$\times$ the minimum measurable torque (\SI{e-8}{\newton\meter}) of our shear rheometer. Continuous line in (b): HB fit to Purell data.  (c) Time evolution of the normalized filament diameter $D(t)/D_0$, where $D_0$ is the initial filament diameter.}  
	\label{fgr:WHO_purell}
\end{figure}

The WHO 80\% formulation is Newtonian with viscosity $2\times$ that of water, Fig.~\ref{fig:WHO_Purell_visc}. The Purell product  shows HB behaviour, Fig.~\ref{fig:WHO_Purell_stress}, with $\sigma_y = \SI{12.7}{\pascal}$, $k = \SI{9.1}{\pascal\second^{0.43}}$ and $n = 0.43$. There was no measurable $N_1$ in either system over our $\dot\gamma$ range. Filament breakage occurs sharply at $\approx \SI{0.3}{\second}$, Fig.~\ref{fig:Purell_filament}.

\subsubsection{WHO + Carbopol Ultrez 20} \label{sec:carb20Rheo}

\begin{figure}
	\centering
	\subfloat[\centering]{\includegraphics[width=0.255\textwidth]{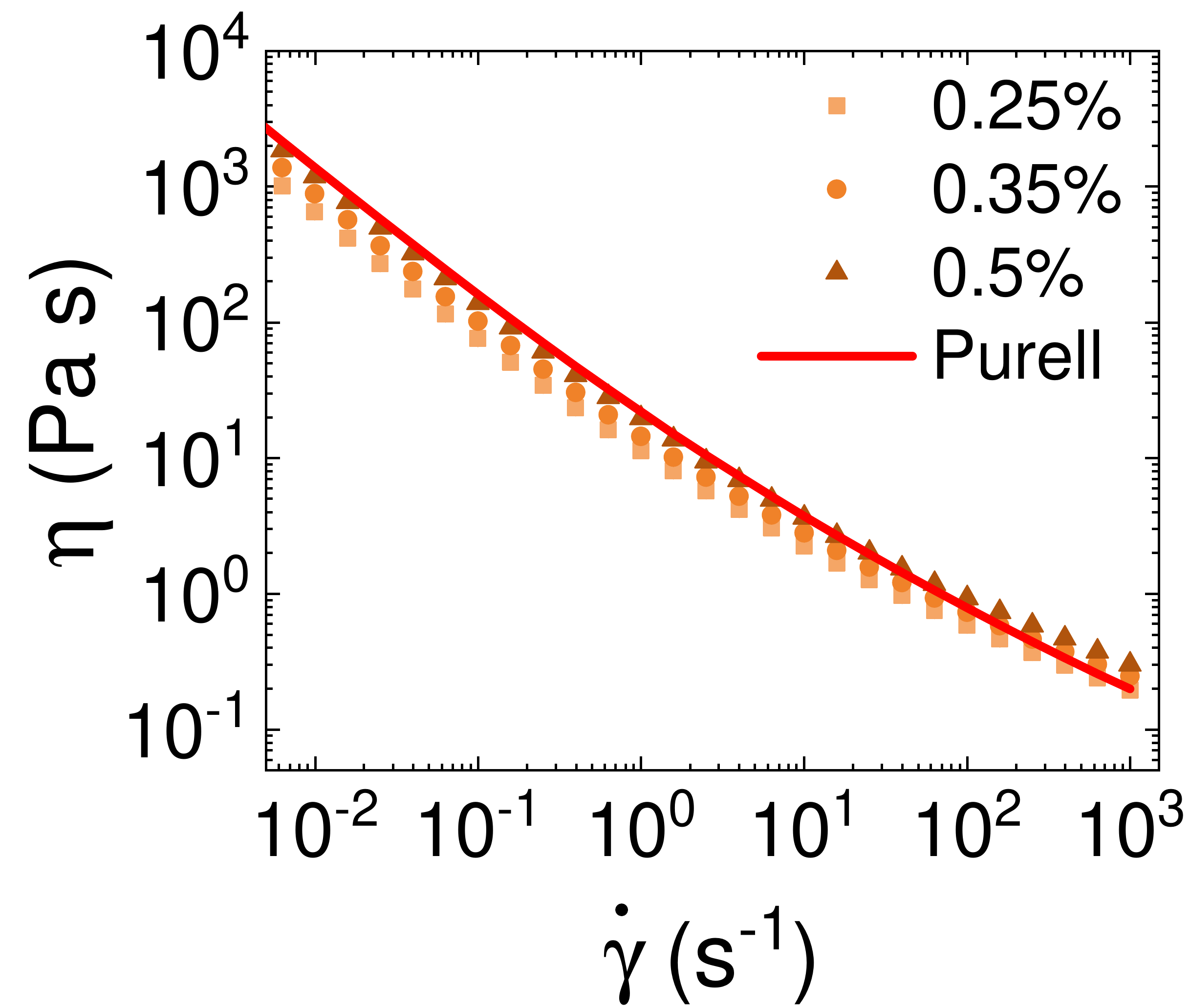}\label{fig:Carbopol_visc}}
	\subfloat[\centering]{\includegraphics[width=0.255\textwidth]{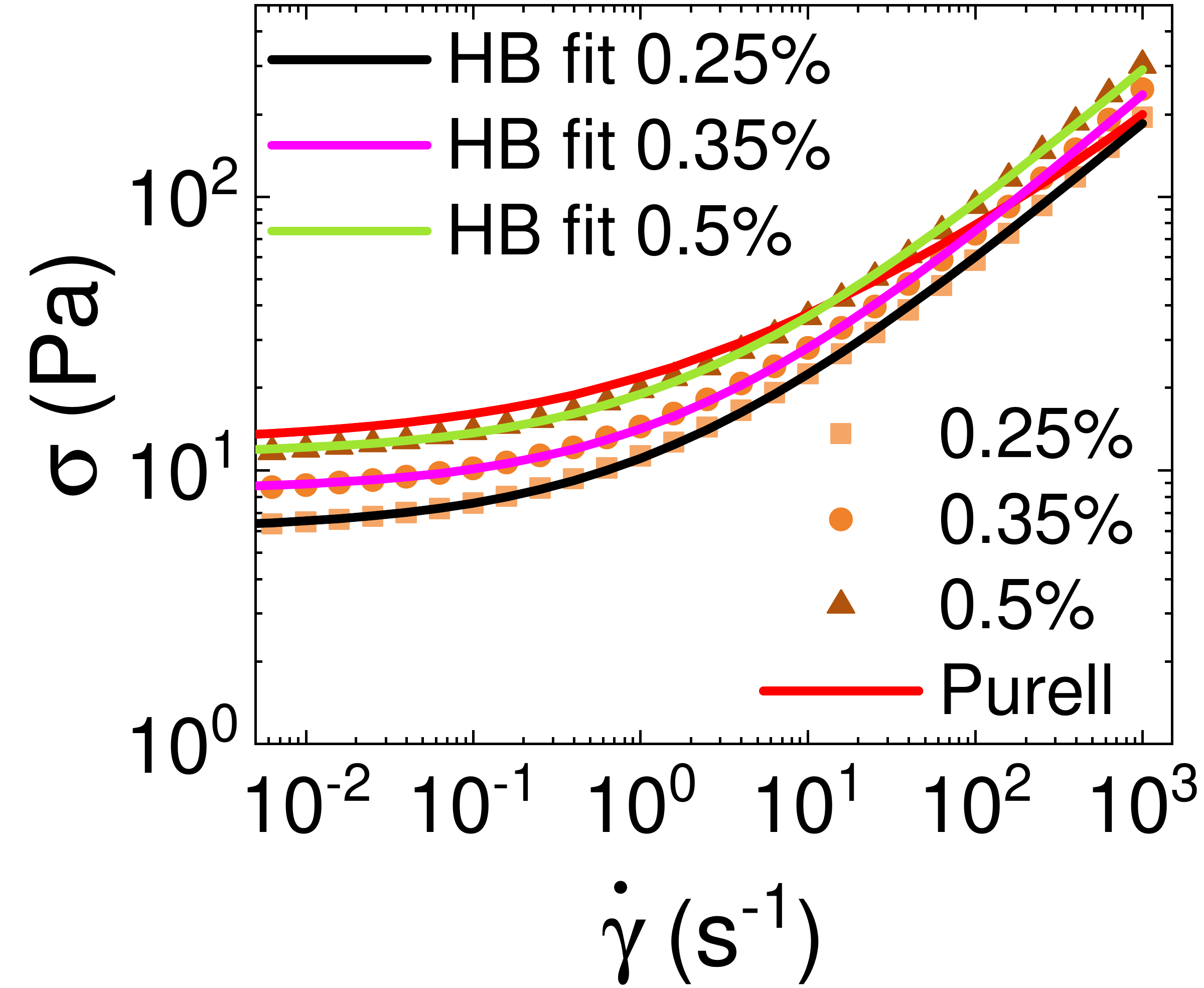}\label{fig:Carbopol_stress}}\\
	\subfloat[\centering]{\includegraphics[width=0.255\textwidth]{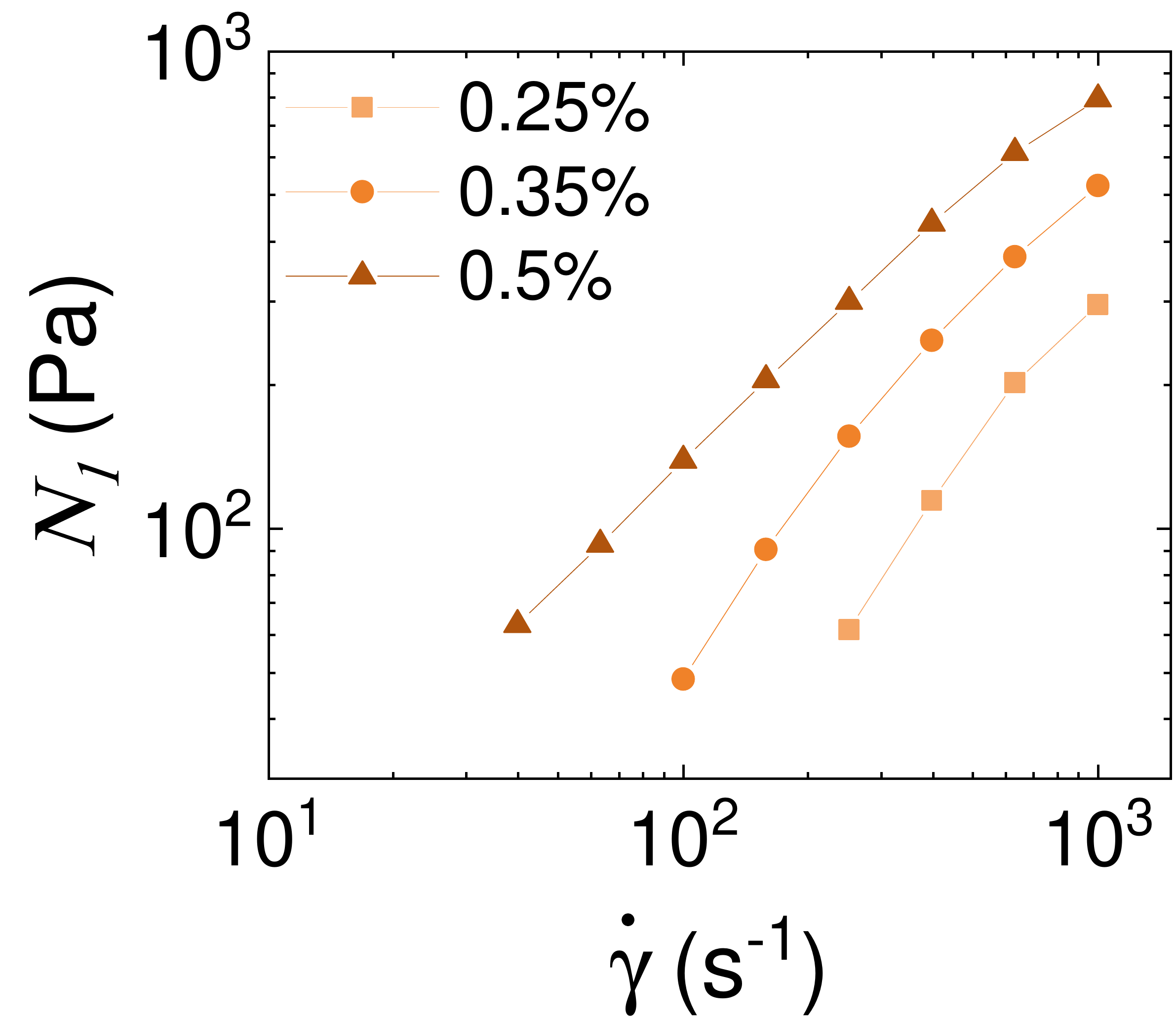}\label{fig:Carbopol_N1}}
	\subfloat[\centering]{\includegraphics[width=0.255\textwidth]{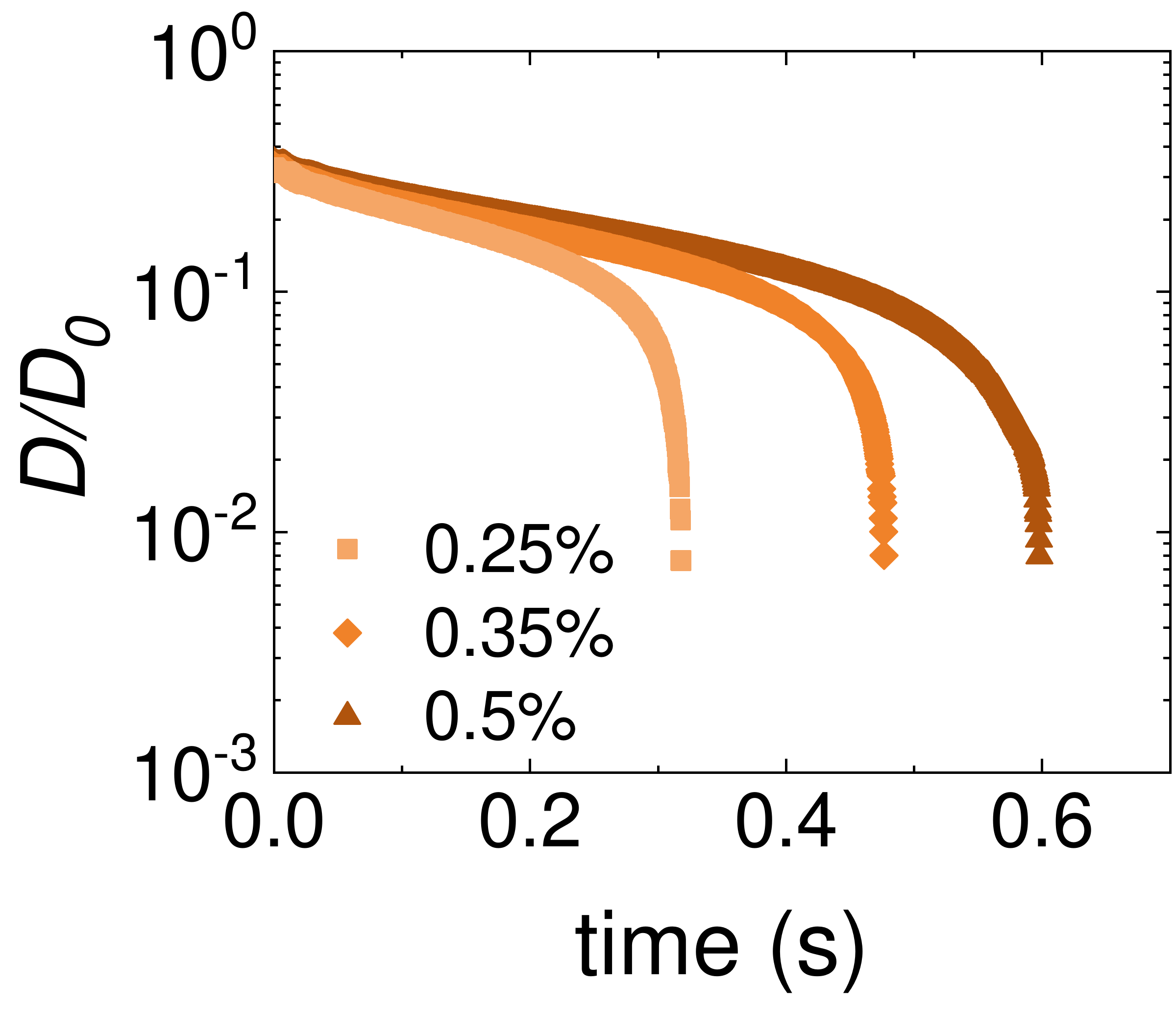}\label{fig:Carbopol_ultrez_filament}}
	\caption{(a) Viscosity, (b) stress vs shear rate and (c) first normal stress difference of WHO formulation thickened with various concentrations of Carbopol Ultrez 20. Red curve gives the Purell data for comparison. Other continuous curves in (b) are HB fits to the data. (d) Time evolution of the normalized filament diameter $D(t)/D_0$.}  
	\label{fgr:Carbopol_ultrez}
\end{figure}

We chose Ultrez~20 to model the Purell gel. Data at 0.5\% indeed closely follow those for the commercial product, Fig.~\ref{fig:Carbopol_visc}-b. Fitting to the HB model gave $(\sigma_y, k, n)$ of $(\SI{6.1}{\pascal}, \SI{5.0}{\pascal\second^{0.52}}, 0.52)$, $(\SI{8.4}{\pascal}, \SI{5.8}{\pascal\second^{0.53}}, 0.53)$ and $(\SI{11.4}{\pascal}, \SI{7.5}{\pascal\second^{0.52}}, 0.52)$ at 0.25\%, 0.35\% and 0.5\%  respectively.  
Interestingly, however, we find measurable $N_1$ in all the samples tested, reaching $\lesssim \SI{e3}{\pascal}$ at the highest $\dot\gamma$ for 0.5\%, Fig.~\ref{fig:Carbopol_N1}. Filament breakage occurs sharply at $\approx 0.3, 0.5$ and \SI{0.6}{\second} with increasing concentration, Fig.~\ref{fig:Carbopol_ultrez_filament}.

\subsubsection{WHO + Carbopol 974P} \label{sec:carbRheo}

\begin{figure}[H]
	\centering
	\subfloat[\centering]{\includegraphics[width=0.255\textwidth]{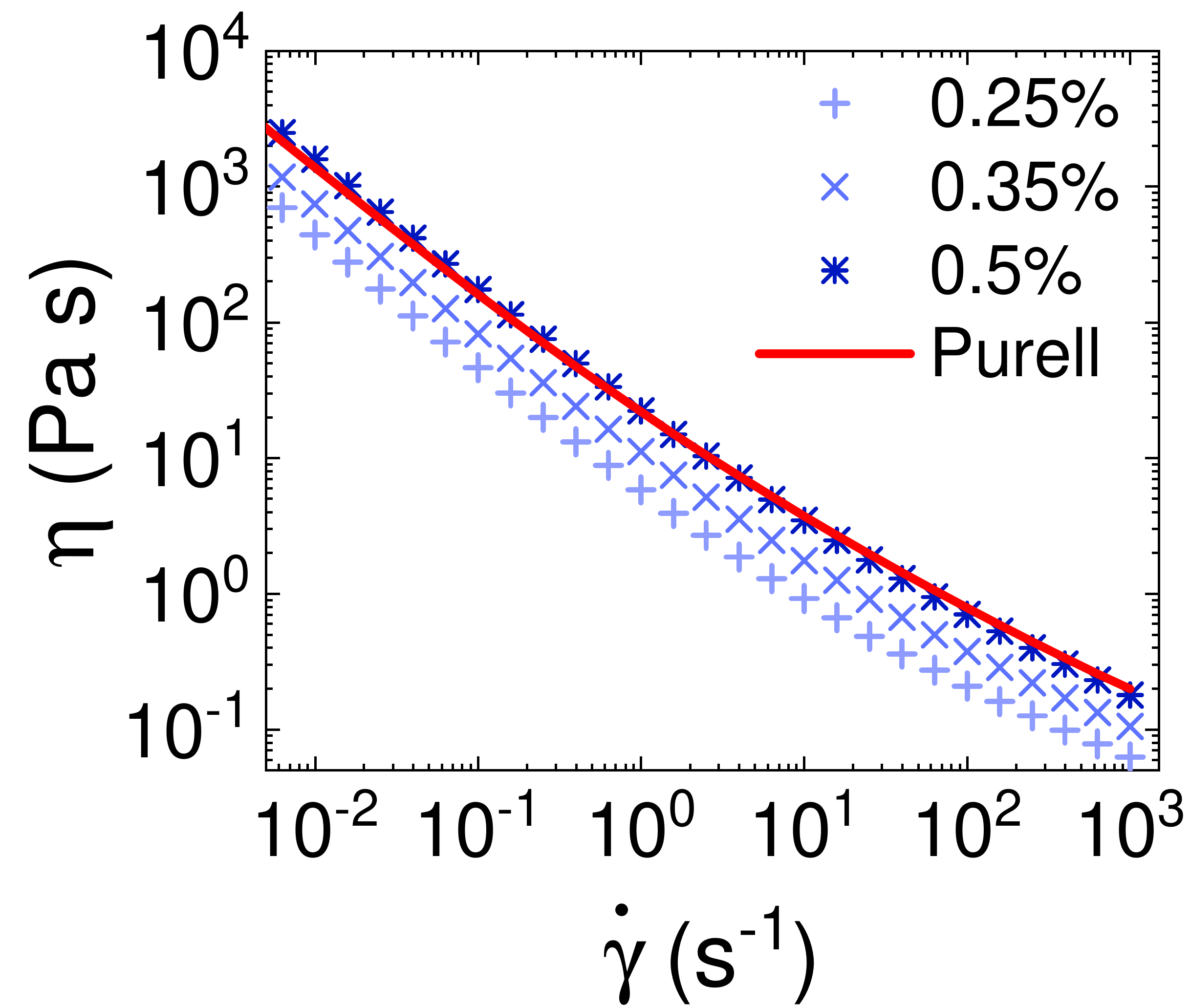}\label{fig:Carbopol_974P_visc}}
	\subfloat[\centering]{\includegraphics[width=0.255\textwidth]{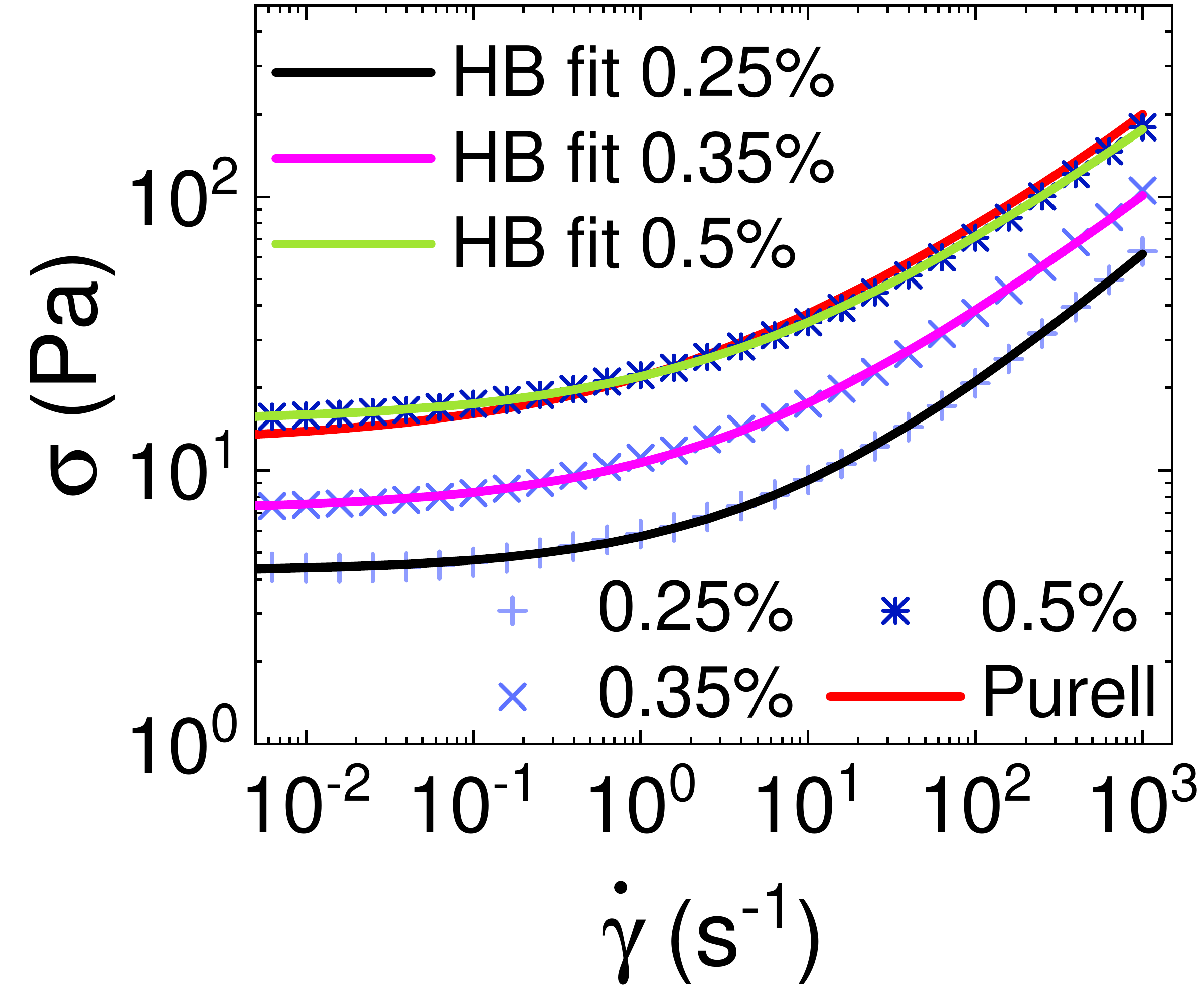}\label{fig:Carbopol_974P_stress}}\\
	\subfloat[\centering]{\includegraphics[width=0.255\textwidth]{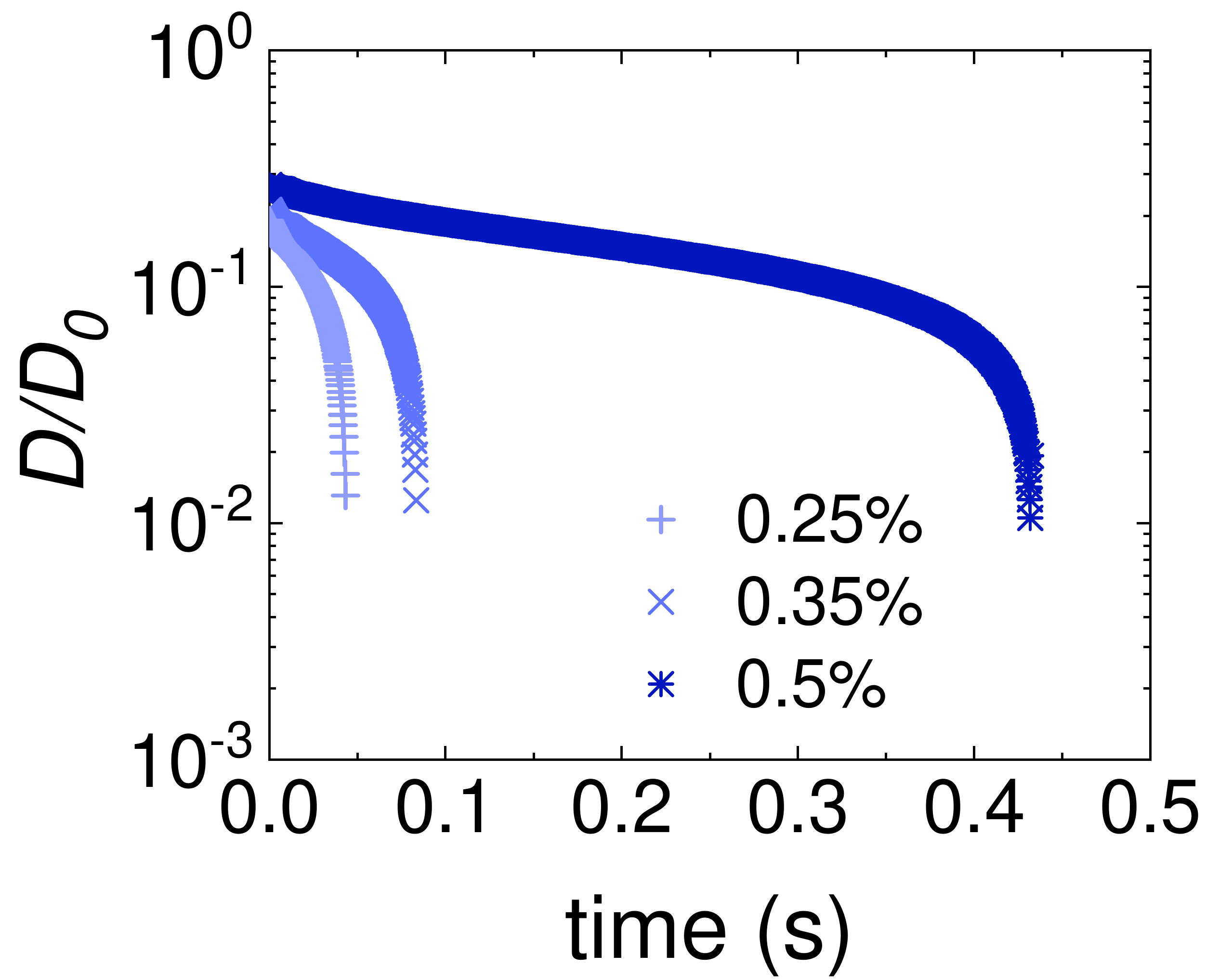}\label{fig:Carbopol_974P_filament}}
	\caption{(a) Viscosity and (b) stress vs shear rate of WHO formulation thickened with various concentrations of Carbopol 974P. Red curve gives the Purell data for comparison. Other continuous curves in (b) are HB fits to the data. (c) Time evolution of the normalized filament diameter $D(t)/D_0$.}
	\label{fgr:Carbopol_974P}
\end{figure}

Our data for WHO + Carbopol 974P at 0.25\%, 0.35\% and 0.5\% polymer concentration, Fig.~\ref{fig:Carbopol_974P_visc}-b, can again be fitted to the HB model with $(\sigma_y, k, n)$ of $(\SI{4.3}{\pascal}, \SI{1.4}{\pascal\second^{0.53}}, 0.53)$, $(\SI{7.1}{\pascal}, \SI{3.5}{\pascal\second^{0.48}}, 0.48)$ and $(\SI{15.1}{\pascal}, \SI{6.9}{\pascal\second^{0.46}}, 0.46)$ respectively.
Data for the 0.5\% gel closely follows that for the Purell hand gel. Now, however, 
we find no measurable $N_1$ in any of these samples over our $\dot\gamma$ range. Filament breakage occurs sharply at $\approx 0.05, 0.1$ and \SI{0.45}{\second} with increasing concentration, Fig.~\ref{fig:Carbopol_974P_filament}. 

\subsubsection{WHO + Sepimax Zen}

\begin{figure}[ht]
	\centering
	\subfloat[\centering]{\includegraphics[width=0.255\textwidth]{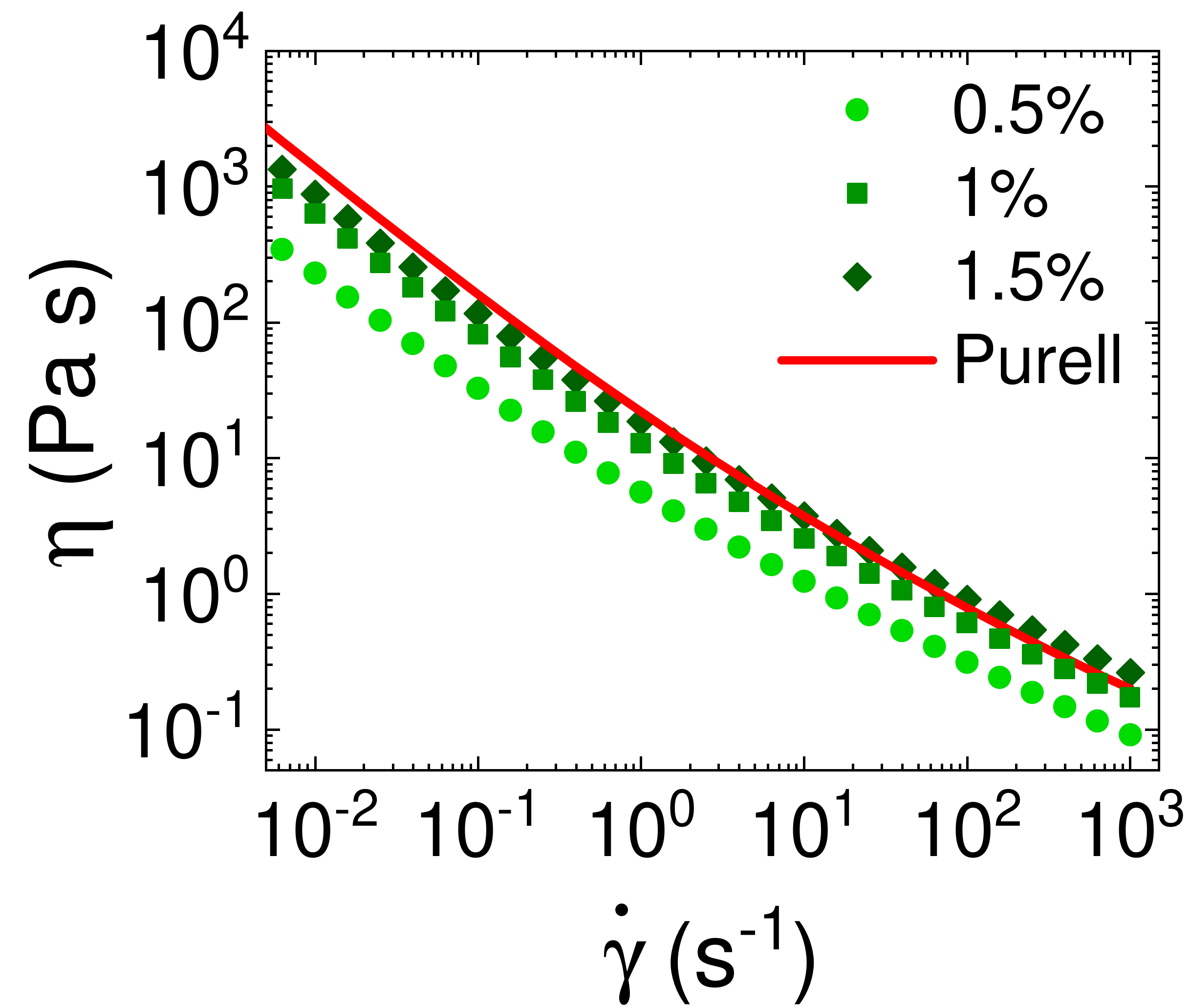}\label{fig:Sepimax_visc}}
	\subfloat[\centering]{\includegraphics[width=0.255\textwidth]{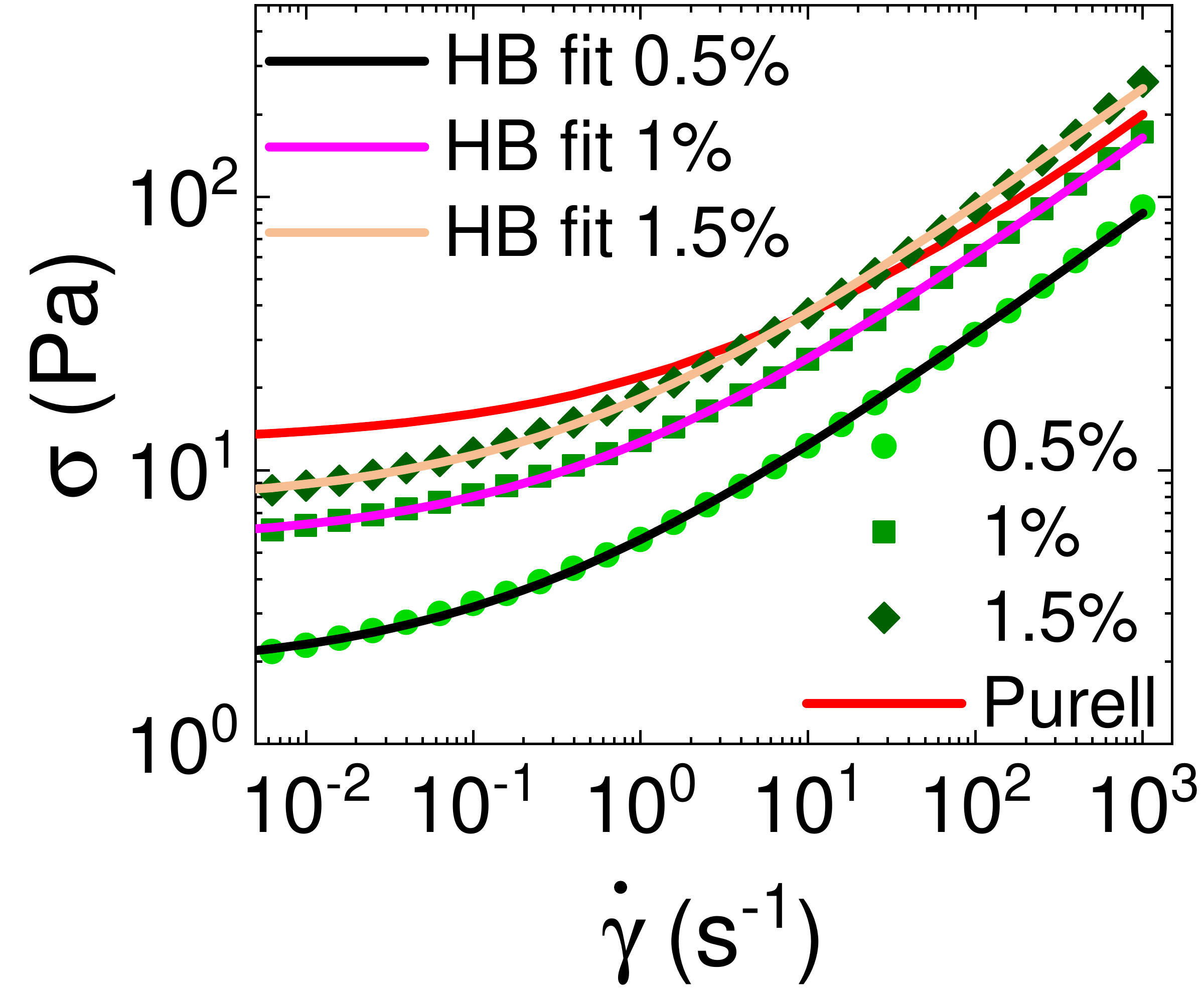}\label{fig:Sepimax_stress}}\\
	\subfloat[\centering]{\includegraphics[width=0.255\textwidth]{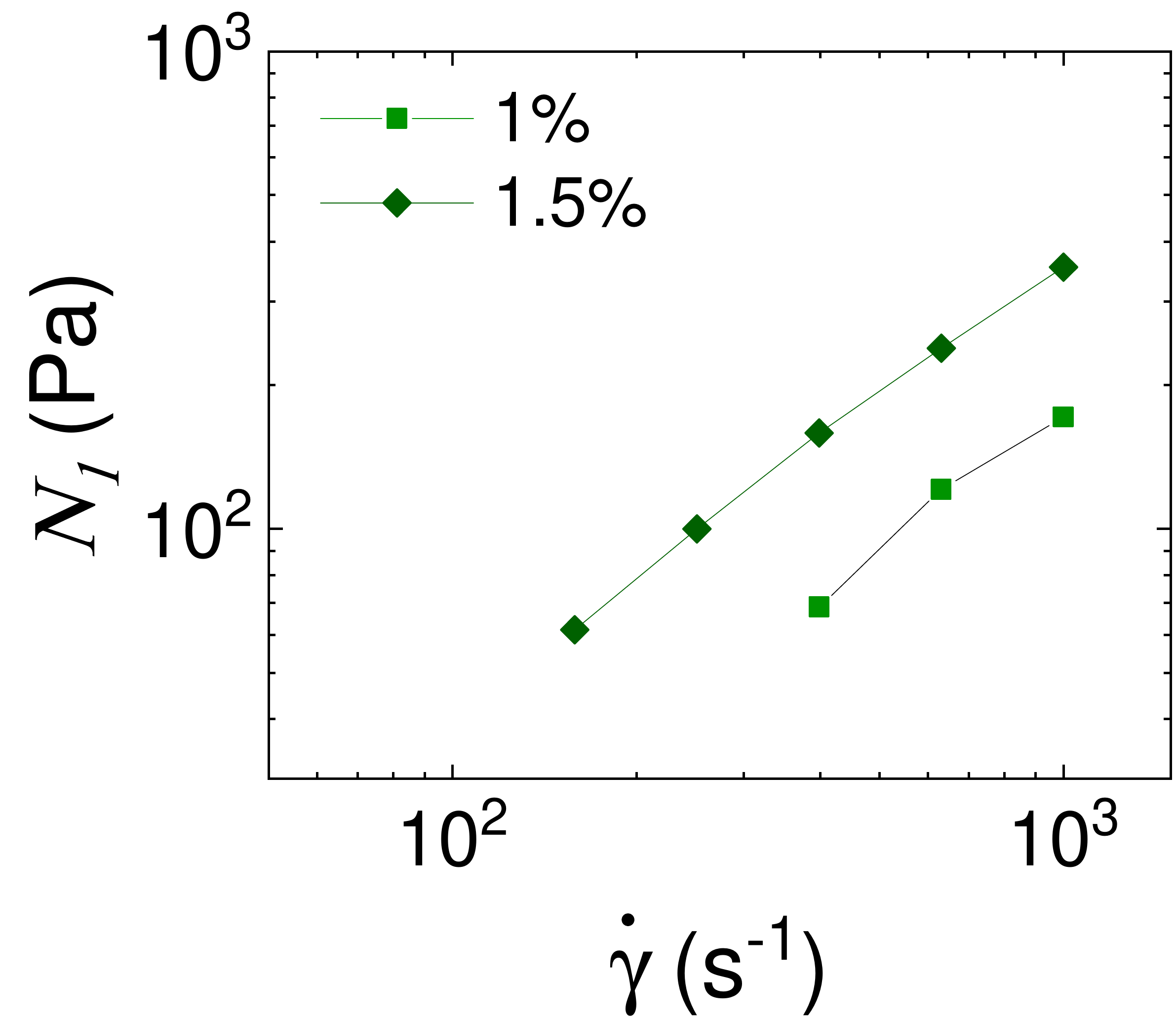}\label{fig:Sepimax_normal}}
	\subfloat[\centering]{\includegraphics[width=0.26\textwidth]{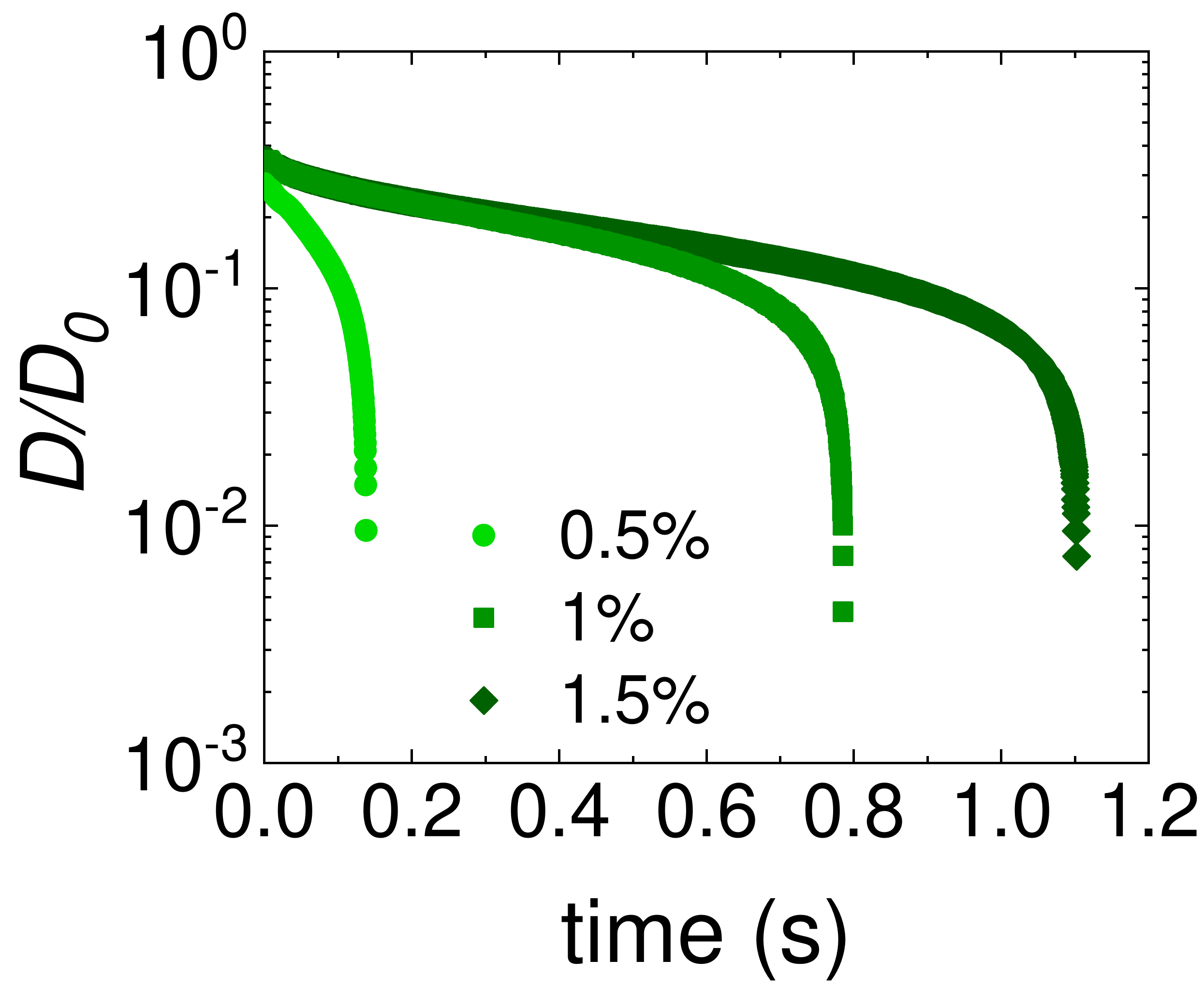}\label{fig:Sepimax_filament}}
	\caption{(a) Viscosity, (b) stress vs shear rate and (c) first normal stress difference of WHO formulation thickened with various concentrations of Sepimax Zen. Red curve gives the Purell data for comparison. Other continuous curves in (b) are HB fits to the data. (d) Time evolution of the normalized filament diameter $D(t)/D_0$.}
	\label{fgr:Sepimax}
\end{figure}

Mixing the WHO formulation  with 0.5\%, 1\% and 1.5\% Sepimax Zen gave gels showing HB behaviour, Fig.~\ref{fgr:Sepimax}a-b, with $(\sigma_y, k, n) = (\SI{1.9}{\pascal}, \SI{3.7}{\pascal\second^{0.45}}, 0.45)$, $(\SI{5.4}{\pascal}, \SI{7.2}{\pascal\second^{0.45}}, 0.45)$ and $(\SI{7.5}{\pascal}, \SI{10.8}{\pascal\second^{0.45}}, 0.45)$ respectively. Thus, $\gtrsim 1.5\%$ of Sepimax is needed to mimic the rheology of the Purell product, while this was achieved with only 0.5\% of Carbopol Ultrez 20 and 974P. This is likely because Carbopols ($\sim 10^9$~Da) have higher molecular weight than Sepimax ($\sim \SI{e5}{\dalton}$).

There was measurable $N_1$ in 1\% and 1.5\% Sepimax gels for $\dot\gamma>\SI{100}{\second^{-1}}$, Fig.~\ref{fig:Sepimax_normal}, and we found abrupt filament breakage at $\approx 0.15, 0.8$ and \SI{1.1}{\second} respectively, Fig.~\ref{fig:Sepimax_filament}.

\subsubsection{WHO + Jaguar HP 120 COS}

\begin{figure}[ht]
	\centering
	\subfloat[\centering]{\includegraphics[width=0.255\textwidth]{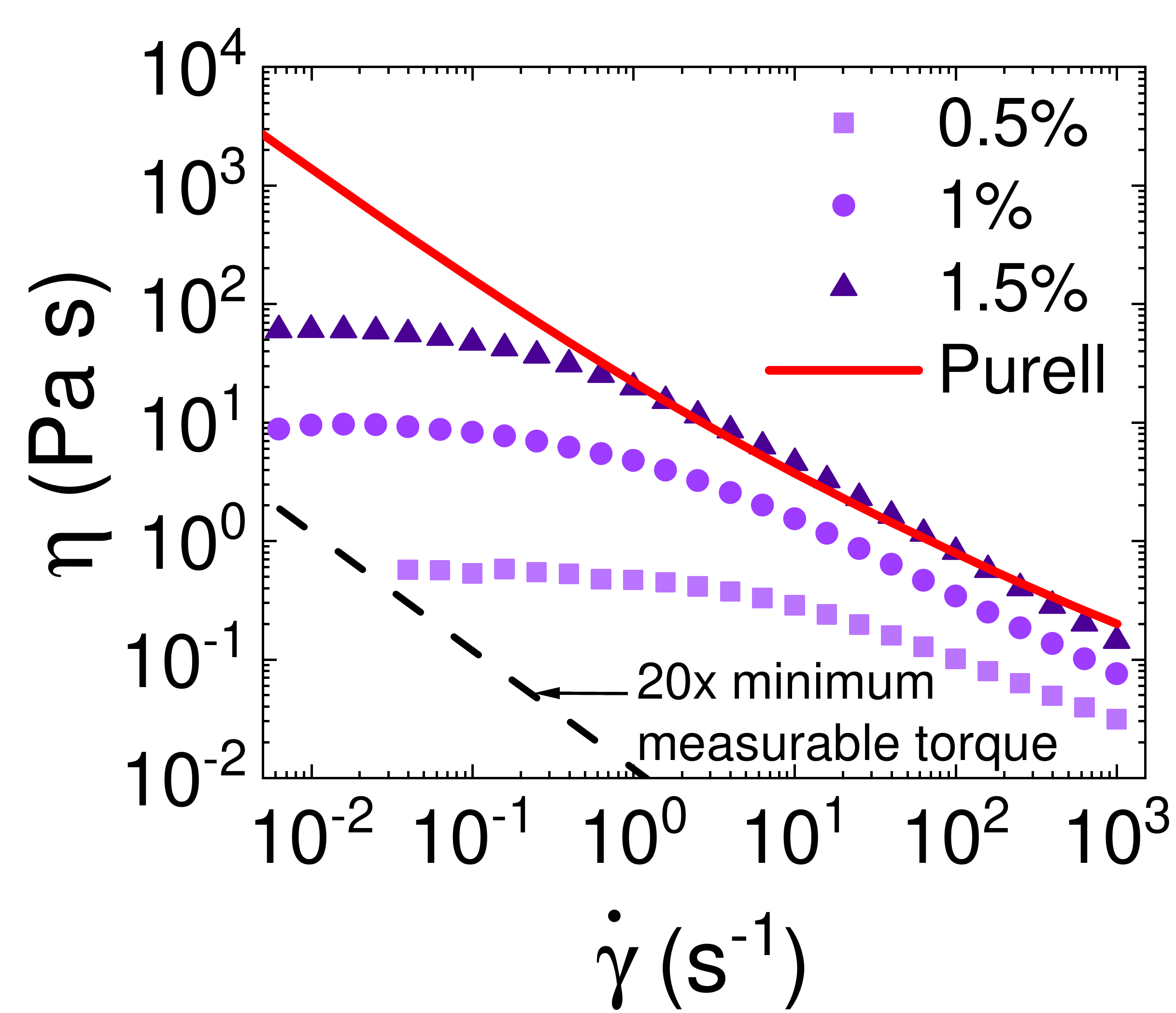}\label{fig:Jaguar_visc}}
	\subfloat[\centering]{\includegraphics[width=0.255\textwidth]{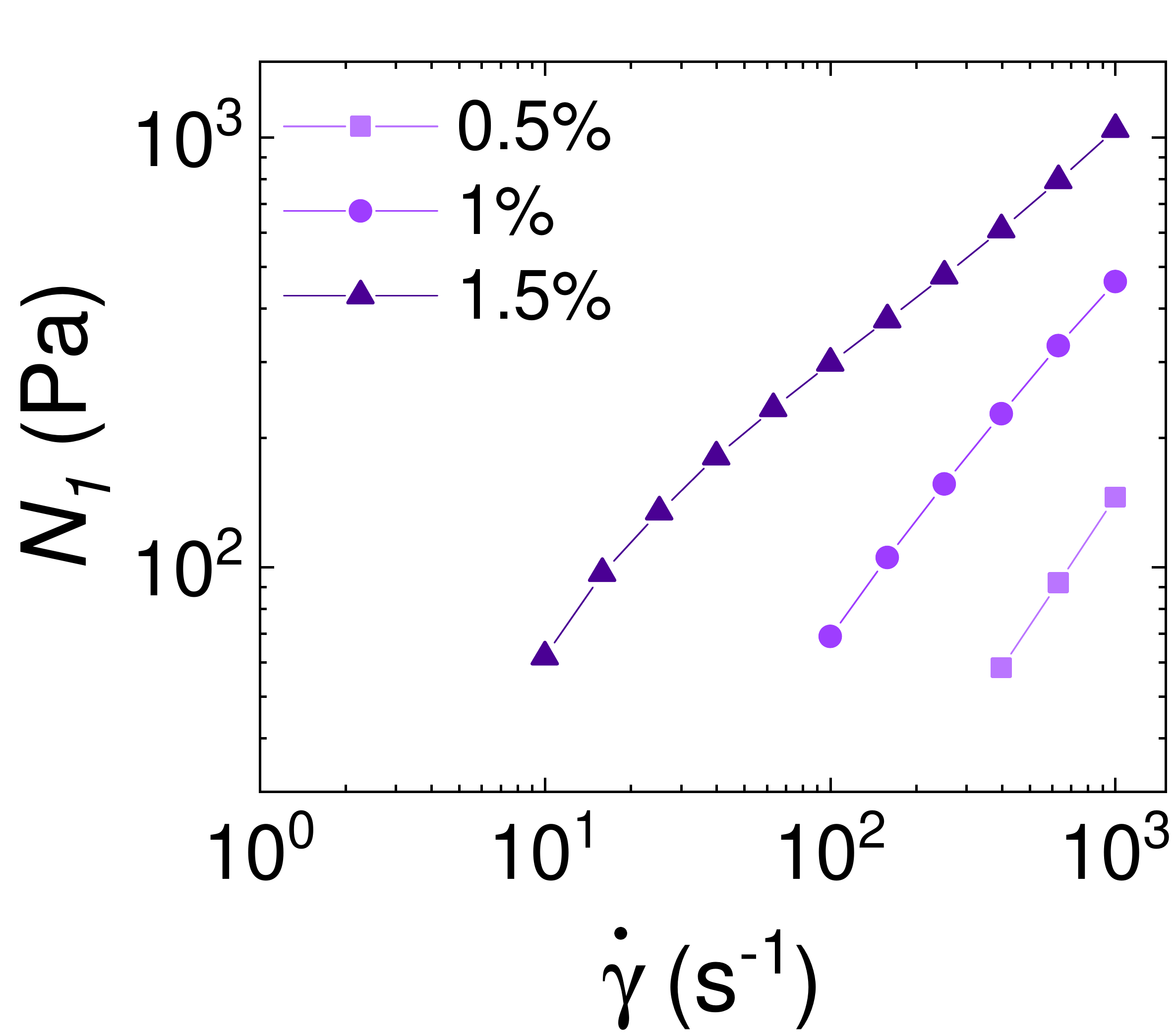}\label{fig:Jaguar_N1}}\\  
	\subfloat[\centering]{\includegraphics[width=0.255\textwidth]{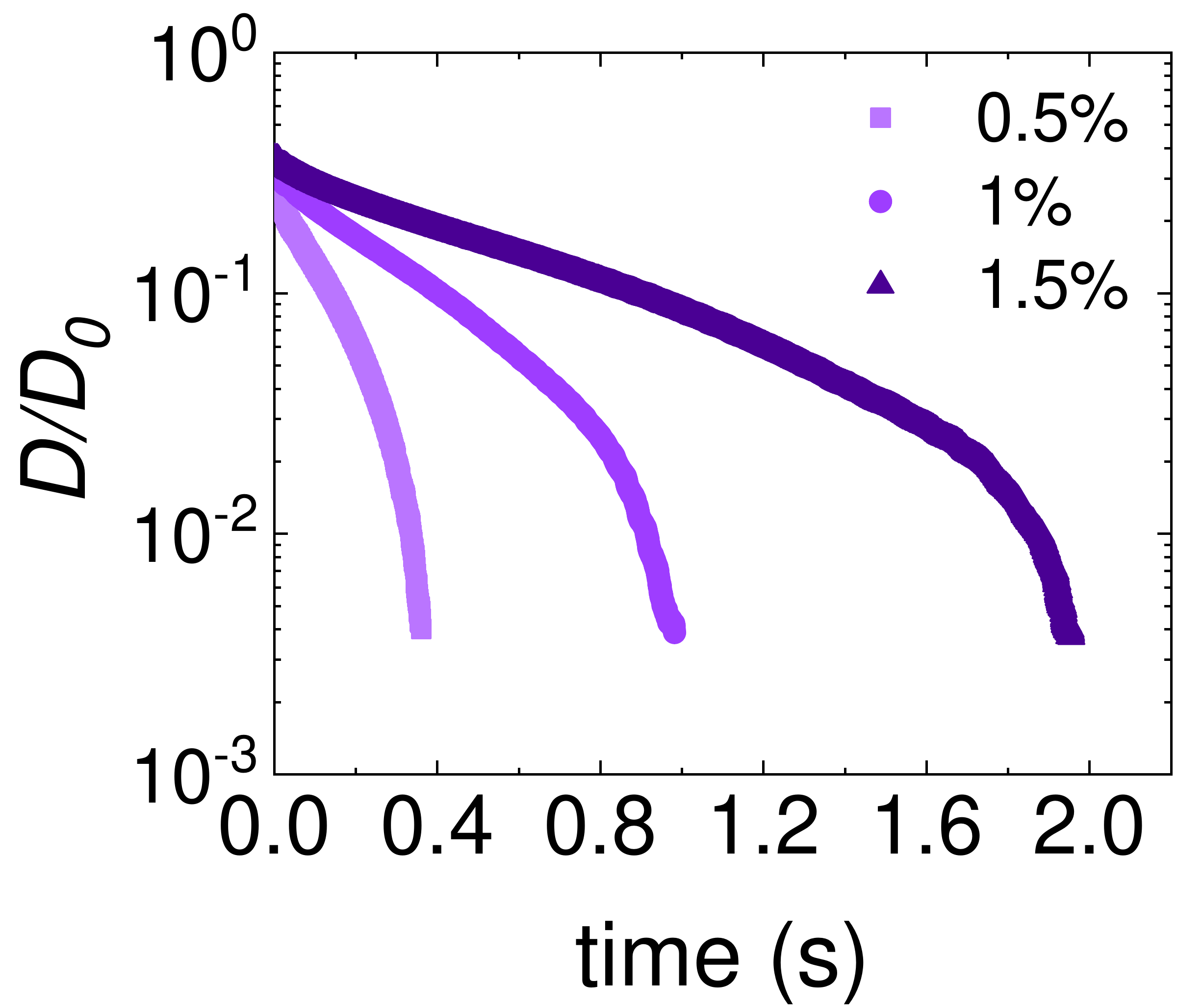}\label{fig:Jaguar_extension}}
	\caption{(a) Viscosity, (b) first normal stress difference of solutions prepared with various concentrations of Jaguar HP 120 COS in the WHO formulation. In (a) the red curve gives the Purell data and the dashed line gives the lowest measurable shear viscosity corresponding to 20$\times$ the minimum measurable torque (\SI{e-8}{\newton\meter}) of our shear rheometer. (c) Time evolution of the normalized filament diameter $D(t)/D_0$.}
	\label{fgr:Jaguar}
\end{figure}

Carbopol Ultrez 20, Carbopol 974P and Sepimax are all microgel particles. By contrast, Jaguar HP 120 COS is a linear polymer. We therefore do not expect, and do not find, a yield stress in the WHO formulation thickened with this material. Instead, the solution shear thins from a finite viscosity at $\dot\gamma \to 0$ through the range of concentrations studied (0.5\%, 1\% and 1.5\%), Fig.~\ref{fig:Jaguar_visc}.

A measurable $N_1$ is found at high $\dot\gamma$, Fig.~\ref{fig:Jaguar_N1}. Filament breakage is less abrupt than is seen in the other samples we have encountered so far, and occurs at $\approx  0.35, 1$ and \SI{2}{\second} with increasing concentration, Fig.~\ref{fig:Jaguar_extension}. 

\section{Discussion}

\subsection{Design criteria}

In Sec.~\ref{sec:design} we proposed a number of rheological design criteria for ABHRs to achieve handleability and give a number of desirable sensory properties such as `smoothness' and `non-stickiness'. We now evaluate the formulations we have characterised against these criteria, Table~\ref{tab:compare}. 

\begin{table*}
	\caption{Comparing ABHR formulations against our design criteria. \textcolor{ForestGreen}{green} = yes, \textcolor{blue}{blue} = marginal, \textcolor{red}{red} = no, \textcolor{red}{$\times$} = not measurable, -- = not applicable.} \label{tab:compare}.
	\begin{adjustbox}{width=1\textwidth}
		\begin{tabular}{|l|c|c|c|c|}
			\hline
			\multicolumn{1}{|c|}{Sample} & \multicolumn{1}{c|}{\begin{tabular}[c]{@{}c@{}}Low run off: \\ $\sigma_y \gtrsim \SI{10}{\pascal}$ or $\eta \gtrsim \SI{1}{\pascal\second}$ \\ at $\dot\gamma \lesssim \SI{0.1}{\per\second}$\end{tabular}} & \multicolumn{1}{c|}{\begin{tabular}[c]{@{}c@{}}Spreadability: \\ shear thins to $\eta \approx \SI{e-1}{\pascal\second}$ \\ at $\dot\gamma \sim \SI{e3}{\per\second}$\end{tabular}} & \multicolumn{1}{c|}{\begin{tabular}[c]{@{}c@{}}Smoothness: \\ $N_1$ at $\dot\gamma \sim \SI{e3}{\per\second}$ \end{tabular}} & \multicolumn{1}{c|}{\begin{tabular}[c]{@{}c@{}}Non-sticky: \\ $\tau_{\rm b} \lesssim \SI{1}{\second}$\end{tabular}} \\  \hline\hline 
			WHO & \textcolor{red}{$\times$} & --  & -- &  -- \\ \hline
			Purell & \textcolor{ForestGreen}{$\sigma_y = \SI{12.7}{\pascal}$} and \textcolor{ForestGreen}{$\eta = \SI{161}{\pascal\second}$}  &  \textcolor{ForestGreen}{$\eta = \SI{0.2}{\pascal\second}$}  & \textcolor{red}{$\times$} &  \textcolor{ForestGreen}{$\tau_{\rm b} = \SI{0.3}{\second}$}  \\ \hline
			0.25\% Carbopol Ultrez 20 & \textcolor{red}{$\sigma_y = \SI{6.1}{\pascal}$} and \textcolor{ForestGreen}{$\eta = \SI{76}{\pascal\second}$}  &  \textcolor{ForestGreen}{$\eta = \SI{0.2}{\pascal\second}$}  & \textcolor{ForestGreen}{$N_1= \SI{295}{\pascal}$}  &  \textcolor{ForestGreen}{$\tau_{\rm b} = \SI{0.3}{\second}$} \\ \hline
			0.35\% Carbopol  Ultrez 20& \textcolor{ForestGreen}{$\sigma_y = \SI{8.4}{\pascal}$} and \textcolor{ForestGreen}{$\eta = \SI{102}{\pascal\second}$}   & \textcolor{ForestGreen}{$\eta = \SI{0.25}{\pascal\second}$}  &  \textcolor{ForestGreen}{$N_1= \SI{524}{\pascal}$} & \textcolor{ForestGreen}{$\tau_{\rm b} = \SI{0.5}{\second}$} \\ \hline
			0.5\% Carbopol   Ultrez 20& \textcolor{ForestGreen}{$\sigma_y = \SI{11.4}{\pascal}$} and \textcolor{ForestGreen}{$\eta = \SI{138}{\pascal\second}$}  & \textcolor{ForestGreen}{$\eta = \SI{0.3}{\pascal\second}$} & \textcolor{ForestGreen}{$N_1= \SI{795}{\pascal}$} & \textcolor{ForestGreen}{$\tau_{\rm b} = \SI{0.6}{\second}$} \\ \hline
			0.25\% Carbopol 974P & \textcolor{red}{$\sigma_y = \SI{4.3}{\pascal}$} and \textcolor{ForestGreen}{$\eta = \SI{46}{\pascal\second}$}  & \textcolor{blue}{$\eta = \SI{0.06}{\pascal\second}$}   & \textcolor{red}{$\times$} & \textcolor{ForestGreen}{$\tau_{\rm b} = \SI{0.05}{\second}$} \\ \hline
			0.35\% Carbopol  974P& \textcolor{blue}{$\sigma_y = \SI{7.1}{\pascal}$} and \textcolor{ForestGreen}{$\eta = \SI{82}{\pascal\second}$}\  & \textcolor{ForestGreen}{$\eta = \SI{0.1}{\pascal\second}$}  & \textcolor{red}{$\times$} & \textcolor{ForestGreen}{$\tau_{\rm b} = \SI{0.1}{\second}$} \\ \hline
			0.5\% Carbopol  974P & \textcolor{ForestGreen}{$\sigma_y = \SI{15.1}{\pascal}$} and \textcolor{ForestGreen}{$\eta = \SI{175}{\pascal\second}$}  & \textcolor{ForestGreen}{$\eta = \SI{0.18}{\pascal\second}$} & \textcolor{red}{$\times$}  & \textcolor{ForestGreen}{$\tau_{\rm b} = \SI{0.45}{\second}$} \\ \hline
			0.5\% Sepimax Zen    & \textcolor{red}{$\sigma_y = \SI{1.9}{\pascal}$} and \textcolor{ForestGreen}{$\eta = \SI{33}{\pascal\second}$}  & \textcolor{ForestGreen}{$\eta = \SI{0.09}{\pascal\second}$}  & \textcolor{red}{$\times$} &  \textcolor{ForestGreen}{$\tau_{\rm b} = \SI{0.15}{\second}$} \\ \hline
			1\% Sepimax  Zen & \textcolor{red}{$\sigma_y = \SI{5.4}{\pascal}$} and \textcolor{ForestGreen}{$\eta = \SI{81}{\pascal\second}$} & \textcolor{ForestGreen}{$\eta = \SI{0.17}{\pascal\second}$} & \textcolor{ForestGreen}{$N_1= \SI{172}{\pascal}$} &   \textcolor{ForestGreen}{$\tau_{\rm b} = \SI{0.8}{\second}$}   \\ \hline
			1.5\% Sepimax Zen  & \textcolor{blue}{$\sigma_y = \SI{7.5}{\pascal}$} and \textcolor{ForestGreen}{$\eta = \SI{116}{\pascal\second}$} & \textcolor{ForestGreen}{$\eta = \SI{0.26}{\pascal\second}$} & \textcolor{ForestGreen}{$N_1= \SI{353}{\pascal}$}  &   \textcolor{blue}{$\tau_{\rm b} = \SI{1.1}{\second}$}    \\ \hline
			0.5\% Jaguar HP 120 COS  & \textcolor{red}{$\eta = \SI{0.53}{\pascal\second}$} & \textcolor{red}{$\eta = \SI{0.03}{\pascal\second}$}  & \textcolor{ForestGreen}{$N_1= \SI{145}{\pascal}$} & \textcolor{ForestGreen}{$\tau_{\rm b} = \SI{0.35}{\second}$}  \\ \hline
			1\% Jaguar HP 120 COS  & \textcolor{ForestGreen}{$\eta = \SI{8}{\pascal\second}$}  & \textcolor{ForestGreen}{$\eta = \SI{0.08}{\pascal\second}$} & \textcolor{ForestGreen}{$N_1= \SI{462}{\pascal}$} & \textcolor{blue}{$\tau_{\rm b} = \SI{1}{\second}$} \\ \hline
			1.5\% Jaguar HP 120 COS & \textcolor{ForestGreen}{$\eta = \SI{47}{\pascal\second}$}  & \textcolor{ForestGreen}{$\eta = \SI{0.14}{\pascal\second}$} & \textcolor{ForestGreen}{$N_1= \SI{1045}{\pascal}$} & \textcolor{red}{$\tau_{\rm b} = \SI{2}{\second}$}\\ \hline
		\end{tabular}
	\end{adjustbox}
\end{table*}

All but the most dilute Jaguar HP 120 COS `minimal formulations' should significantly alleviate run off, and show the required degree of shear thinning (to $\approx \SI{e-1}{\pascal\second}$ at $\dot\gamma \sim \SI{e3}{\per\second}$) to give a desired degree of `spreadability'. 

Interestingly, all but the Purell product and WHO + Carbopol 974P formulation show some measurable $N_1$ at high $\dot\gamma$, which may confer the  sensory property of smoothness during final rubdown \citep{Tamura2013}.  

Where the formulations may show most variability is in the degree of stickiness. We suppose that if filaments formed between a retracting thumb and forefinger take $\lesssim \SI{1}{\second}$ to break, a product will not be perceived as sticky. By this criterion, our formulations range from the acceptable to marginal and unacceptable.

In this context, we also note that the time evolution of filament diameters in all formulations thickened with microgels show a characteristic shape, typified by the data for Ultrez 20, Fig.~\ref{fig:Carbopol_ultrez_filament}. Such abrupt breakage of filaments is also found in other yield-stress materials such as jammed emulsions \citep{Brummer2009}. The data for Jaguar, however, are qualitatively different, and are typical of entangled polymer solution \citep{Arnolds2010}. Here, elastic filaments are created and the filament thinning right before breakup is less abrupt. It will be interesting to further investigate the potentially different sensory perception offered during rubbing by these two kinds of behaviour.

\subsection{Film stability}

We have already seen that most of our thickened formulations shear thin to the requisite degree to confer the benefit of `spreadability'. It turns out that the rapidity of shear thinning may also impact on formulation acceptability. 

As a topical formulation such as an ABHR is rubbed down, it acts as a  lubricant between the finger and the skin of the site of application. During this process, the system remains in the so-called hydrodynamic lubrication (HL) regime until the stresses involved become sufficient to deform the asperities on the skin surface, where upon the system progressively transitions into the so-called elastohydrodynamic (EHL) regime. 

The full scenario is complex and still under investigation \citep{Adams2007,Persson2013}. However, the fundamentals of HL between hard surfaces are well understood \citep{Hall1978,HSJ2004}. Nevertheless, it is only recently that the {\it stability} of the load-bearing fluid film in HL has been addressed \citep{Warren2017}. The result for the latter is of some interest in the present context.

It is well known that there is a one-to-one relationship between the load $W$ borne by a HL fluid film and its minimum thickness $h_0(W)$. Stability is concerned with the question: if the film thickness is  momentarily perturbed away from its equilibrium value $h_0(W)$, would the system spontaneously return to this thickness, or would the perturbation grow? For the case of a sphere rubbing against an infinite plane, Warren shows that a Newtonian HL film is stable. On the other hand, for a shear thinning fluid whose high-shear viscosity scales as $\eta \sim \dot\gamma^{-\alpha}$, stability of the HL film between a sphere and a flat requires $\alpha < 0.5$. On the other hand, Warren's analysis for a cylinder sliding on an infinite plane returns the stability criterion of $\alpha < 1$. 

%\footnote{Note that $\alpha$ relates to the Herschel-Bulkley index via $\alpha = 1 - n$.}

\begin{figure}
	\centering
	\includegraphics[width=0.45\textwidth]{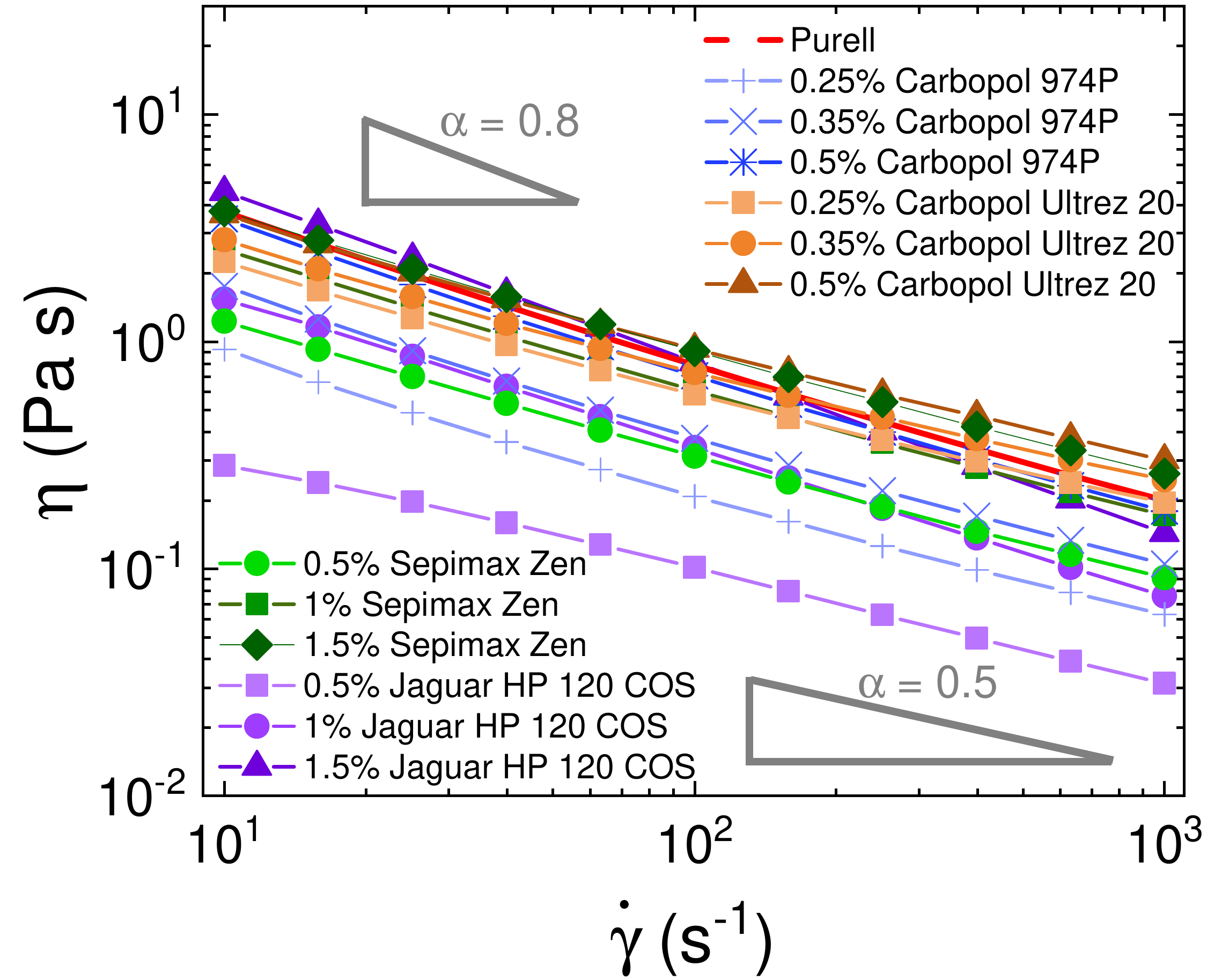}
	\caption{Log-log plot of the high-shear region of the flow curve, $\eta(\dot\gamma)$ of our Carbopol, Sepimax and Jaguar formulations. We find power law shear thinning in all cases: $\eta \sim \dot\gamma^{-\alpha}$. The shear thinning index is $\alpha < 1$ for all formulations tested.}
	\label{fig:index}
\end{figure}

Interestingly, the high-shear viscosity of all of the non-Newtonian samples we tested show $\eta \sim \dot\gamma^{-\alpha}$ with $0.5 \lesssim \alpha \lesssim 0.8$, Fig.~\ref{fig:index}, consistent with previous studies of other hydroxypropyl guar gum solutions \citep{Berardi2020,Lapasin1995} and Carbopol solutions \citep{Frisken2006}. Depending on whether cylinder-on-plane or sphere-on-plane is the more appropriate model for finger rubbing, our formulations should either all show film stability or all show film instability. 

We speculate that such stability may be important for an acceptable skin-feel towards the end of the rubbing-in process, so that this matter clearly deserves detailed future study. Here, we simply note that previous work characterising tactile perception has used a cylinder-on-plane geometry 
\citep{Lips2012}, in which all of our formulations should show film stability.

\section{Conclusions}

We started by articulating a number of science-based principles for designing the rheology of an ABHR that avoids rapid run-off and likely to show desirable hand feel. We then measured the rheology of a commercial product and minimal formulations consisting of WHO hand sanitising liquids thickened with different microgels and a linear polymer. The results were discussed in terms of the principles we articulated. Not surprisingly, our results show that the common practice of thickening ABHRs with microgels to give formulations with finite yield stresses should indeed work. However, our results also suggest that it is possible to prevent run off without a yield stress, provided that the low-shear viscosity is high enough. 

A formulation based on entangled polymers that is without yield stress may be easier to manufacture. Interestingly, however, we find that stretched filaments break differently for such a system than formulations thickened by microgel jamming. The different skin-feel of these two kinds of systems, perhaps particularly vis-\`a-vis stickiness, should be investigated in future work. Linear polymers and certain microgel additives can also potentially impart enhanced `smoothness' through a finite $N_1$.  

While both the commercial product (according to publicly available information) and the Ultrez 20-based formulation are thickened with a hydrophobically-modified carbomer with INCI name `Acrylates/C10-30 alkyl aerylate crosspolymer', we find the closest similarity in both the shear and extension rheology with 0.5\% Carbopol 974P, which is not hydrophobically-modified. This potentially suggests a non-trivial relation between formulation composition and rheological performance, even for seemingly similar additives.

Finally, we have drawn attention to the potential importance of film stability during the rubbing-down of ABHS. This aspect of the triborheology of this and other topical products has not received significant attention to date, and deserves further study.

\section*{Conflicts of interest}
There are no conflicts to declare.

\section*{Acknowledgements}
We thank Patrick Warren (Edinburgh) for illuminating discussions. We thank Solvay for providing a sample of Jaguar HP 120 COS, and Lubrizol for providing a sample of Carbopol Ultrez 20.

\section*{Appendix 1: Run off time}

There are a number of ways to arrive at an order-of-magnitude estimate of the time it takes a \SI{2}{\milli\liter} dose of WHO formulation with a viscosity of $\eta \approx \SI{2}{\milli\pascal\second}$ to run of a $\lesssim \SI{10}{\centi\meter}$ palm inclined at $\alpha \approx \SI{20}{\degree}$, all of which confirms the estimate from experience, namely, $\lesssim \SI{e-1}{\second}$. Perhaps the simplest is to use a textbook result \citep{Batchelor:1967ay} for the typical flow speed of a liquid film (density $\rho$) with volume flux $Q$ per unit width flowing down an incline at angle $\alpha$:
\begin{equation}
U \sim \left( \frac{9Q^2\rho g \sin\alpha}{\eta}\right)^{\frac{1}{3}}.
\end{equation}
If we take a volume flow rate of $\SI{2}{\milli\liter\per\second}$ in a film of width \SI{1}{\centi\meter}, we find $Q = \SI{2}{\centi\meter^2\second^{-1}}$ and $U \sim \SI{0.6}{\meter\per\second}$, giving a run off time of $\lesssim \SI{e-1}{\second}$.

\section*{Appendix 2: Stickiness}

\cite{Wolf2016}~studied the shear and extensional rheology of shear-thinning mixtures of dextran and xanthan (both plant-derived polysaccharides) in water and quantified the correlation between the shear-thinning rheology with consumer perception of, amongst other things, `stickiness on lips' and `stickiness in mouth'. The authors also characterised their samples using capillary rheology. They reported via a table of values that stickiness perception correlates well with increasing extensional viscosity, but did not incorporate this quantitatively in their perception model. 

\begin{figure}[ht]
	\centering
	\includegraphics[width=0.45\textwidth]{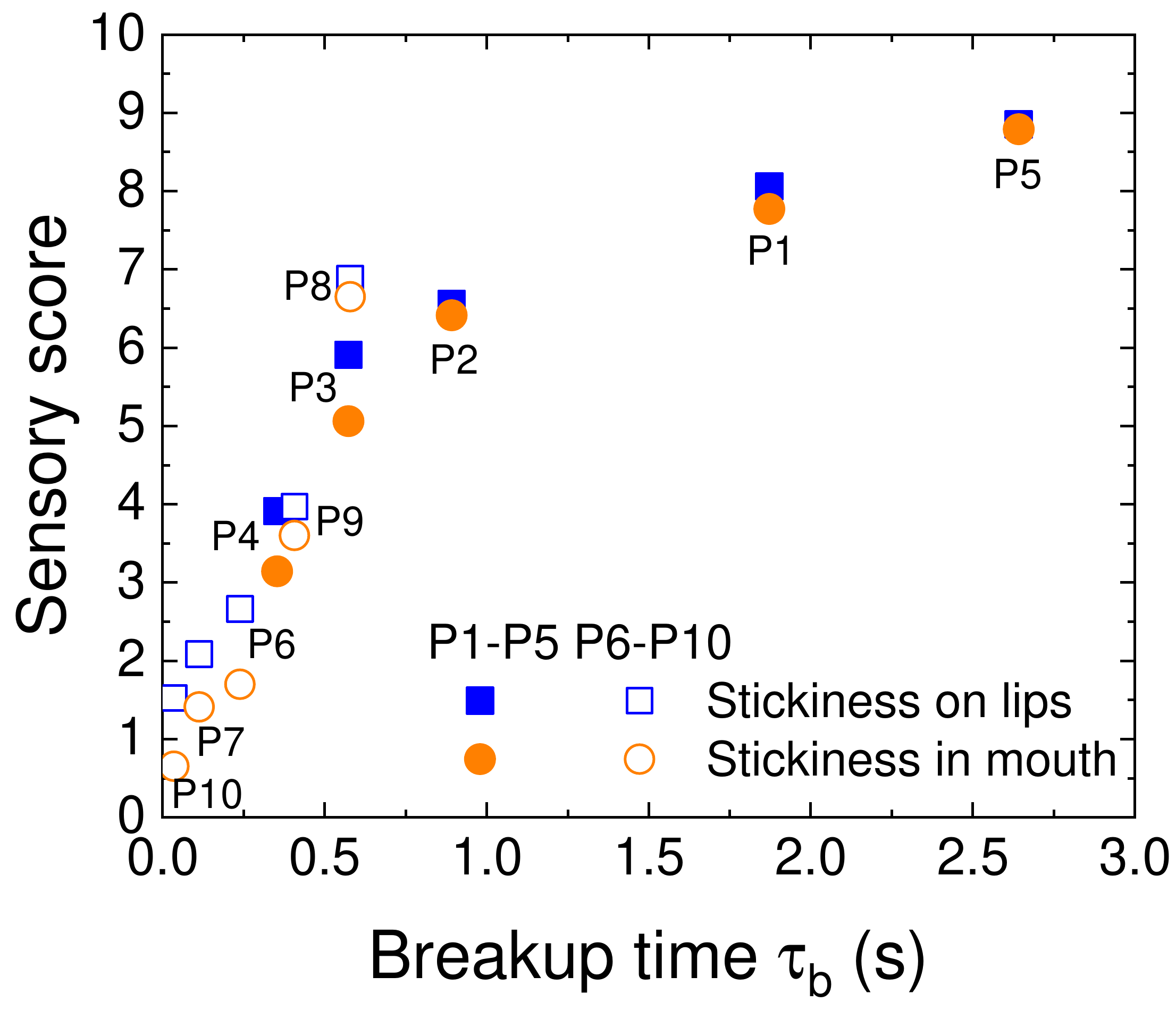}
	\caption{The dependence of consumer scores of perceived on-lips and in-mouth stickiness plotted against the measured capillary breakup time of different samples of aqueous xanthan-dextran mixtures; data taken from \cite{Wolf2016}. Solid symbols correspond to samples P1-P5  which have similar low-shear viscosities, while open symbols correspond to samples P6-P10 which have almost identical high-shear viscosities. }
	\label{fig:sensory_score}
\end{figure}

Usefully, He et al.~also reported filament breaking times as supplementary material. We plot their consumer panel scores of on-lips and in-mouth stickiness against breakup time, $\tau_{\rm b}$ in Fig~\ref{fig:sensory_score}. It is clear that longer breakup times correlate almost perfectly with increased perception of `stickiness' on both types of epithelia. 

Interestingly, the dependence of stickiness on $\tau_{\rm b}$ is highly non-linear: there appears to be two regimes in the data. The perceived stickiness score decreases slowly with breakup time down to $\tau_{\rm b} \approx \SI{1}{\second}$, whereupon the perceived stickiness decreases rapidly. This suggests that $\tau_{\rm b} \lesssim \SI{e0}{\second}$ is a plausible criterion for `acceptable stickiness'. 

There are other subtleties in the data, e.g., that the two regimes in this plot more or less separate samples with approximately the same low-shear viscosity at $\SI{50}{\per\second}$ (a few hundred \si{\pascal\second}, P1-P5) and samples with approximately the same high-shear viscosity at \SI{e5}{\per\second} ($\approx \SI{0.01}{\pascal\second}$, P6-P10). Discussion of these subtleties are, however, beyond the scope of our work.

%%%END OF MAIN TEXT%%%

%The \balance command can be used to balance the columns on the final page if desired. It should be placed anywhere within the first column of the last page.

%\balance

%If notes are included in your references you can change the title from 'References' to 'Notes and references' using the following command:
%\renewcommand\refname{Notes and references}

\bibliography{Hand_gels} 

\begin{thebibliography}{40}
\providecommand{\natexlab}[1]{#1}
\providecommand{\url}[1]{{#1}}
\providecommand{\urlprefix}{URL }
\providecommand{\doi}[1]{\url{https://doi.org/#1}}
\providecommand{\eprint}[2][]{\url{#2}}
 \bibcommenthead

\bibitem[{Adams et~al(2007)Adams, Briscoe, and Johnson}]{Adams2007}
Adams MJ, Briscoe BJ, Johnson SA (2007) Friction and lubrication of human skin.
  Tribol Lett 26:239--253

\bibitem[{Akay et~al(2012)Akay, Echols, Ding, Dussaud, and Lips}]{Lips2012}
Akay A, Echols B, Ding J, et~al (2012) Measurement of sound, vibration and
  friction between soft materials under light loads. Wear 276-277:61--69

\bibitem[{Arnolds et~al(2010)Arnolds, Buggisch, Sachsenheimer, and
  Willenbacher}]{Arnolds2010}
Arnolds O, Buggisch H, Sachsenheimer D, et~al (2010) Capillary breakup
  extensional rheometry (caber) on semi-dilute and concentrated
  polyethyleneoxide (peo) solutions. Rheol~Acta 49:1207--1217

\bibitem[{Batchelor(1967)}]{Batchelor:1967ay}
Batchelor GK (1967) An introduction to fluid dynamics. Cambridge University
  Press, chapter 5

\bibitem[{Bates et~al(1963)Bates, Paabo, and Robinson}]{Bates1963}
Bates RG, Paabo M, Robinson RA (1963) Interpretation of ph measurements in
  alcohol-water solvents. J Phys Chem 67:1833--1838

\bibitem[{Berardi et~al(2020)Berardi, Perinelli, Merchant, Bisharat, Basheti,
  Bonacucina, Cespi, and Palmieri}]{Berardi2020}
Berardi A, Perinelli DR, Merchant HA, et~al (2020) Hand sanitisers amid
  covid-19: A critical review of alcohol-based products on the market and
  formulation approaches to respond to increasing demand. International Journal
  of Pharmaceutics 584:119431

\bibitem[{Bernard et~al(2010)Bernard, Merat, Braun, and Mallo}]{Sepimax}
Bernard P, Merat E, Braun O, et~al (2010) A new polymer with a {MAX}imum
  resistance to electrolytes. {SOFW} J 136(12):55--58

\bibitem[{Bhattacharjee et~al(2018)Bhattacharjee, Kabb, O'Bryan, Urue{\~n}a,
  Sumerlin, Sawyer, and Angelini}]{Angelini2018}
Bhattacharjee T, Kabb CP, O'Bryan CS, et~al (2018) Polyelectrolyte scaling laws
  for microgel yielding near jamming. Soft Matter 14:1559--1570

\bibitem[{Brady et~al(2017)Brady, Dürig, Lee, and Li}]{Brady2017}
Brady J, Dürig T, Lee P, et~al (2017) Chapter 7 - polymer properties and
  characterization. In: Qiu Y, Chen Y, Zhang GG, et~al (eds) Developing Solid
  Oral Dosage Forms, 2nd edn. Academic Press, Boston, p 181 -- 223

\bibitem[{Cheng et~al(2002)Cheng, Brown, and Prud'homme}]{Cheng2002}
Cheng Y, Brown KM, Prud'homme RK (2002) Characterization and intermolecular
  interactions of hydroxypropyl guar solutions. Biomacromolecules 3:456--461

\bibitem[{Crosby(2021)}]{Crosby2}
Crosby D (2021) Exploring the multifunctionality of a giant commercial
  microgel. PhD thesis, The University of Edinburgh

\bibitem[{Dinic and Sharma(2019)}]{Dinic2019}
Dinic J, Sharma V (2019) Macromolecular relaxation, strain, and extensibility
  determine elastocapillary thinning and extensional viscosity of polymer
  solutions. Proc Natl Acad Sci (USA) 116:8766--8774

\bibitem[{Greenaway et~al(2018)Greenaway, Ormandy, Fellows, and
  Hollowood}]{Greenaway2018}
Greenaway R, Ormandy K, Fellows C, et~al (2018) Impact of hand sanitizer format
  (gel/foam/liquid) and dose amount on its sensory properties and acceptability
  for improving hand hygiene compliance. J Hosp Infect 100:195 -- 201

\bibitem[{Gupta and Natarajan(2017)}]{Gupta2017}
Gupta AK, Natarajan U (2017) Anionic polyelectrolyte poly(acrylic acid) (paa)
  chain shrinkage in water–ethanol solution in presence of li+ and cs+ metal
  ions studied by molecular dynamics simulations. Mol Simul 43:625--637

\bibitem[{Halling(1978)}]{Hall1978}
Halling (1978) Principles of Tribology. Macmillan, London

\bibitem[{Hamrock et~al(2004)Hamrock, Schmid, and Jacobsen}]{HSJ2004}
Hamrock BJ, Schmid SR, Jacobsen BO (2004) Fundamentals of {F}luid {F}ilm
  {L}ubrication. Marcel Dekker, New York

\bibitem[{He et~al(2016)He, Hort, and Wolf}]{Wolf2016}
He Q, Hort J, Wolf B (2016) Predicting sensory perceptions of thickened
  solutions based on rheological analysis. Food Hydrocoll 61:221--232

\bibitem[{Kampf et~al(2013)Kampf, Ruselack, Eggerstedt, Nowak, and
  Bashir}]{Kampf2013}
Kampf G, Ruselack S, Eggerstedt S, et~al (2013) Less and less--influence of
  volume on hand coverage and bactericidal efficacy in hand disinfection. BMC
  Infect Dis 13:472

\bibitem[{Katdare and Chaubal(2006)}]{Katdare2006}
Katdare A, Chaubal M (2006) Excipient Development for Pharmaceutical,
  Biotechnology, and Drug Delivery Systems. CRC Press

\bibitem[{Kim et~al(2002)Kim, Lee, and Kang}]{Kim2002}
Kim HY, Lee HJ, Kang BH (2002) Sliding of liquid drops down an inclined solid
  surface. J Colloid Interface Sci 247:372--380

\bibitem[{Kim et~al(2003)Kim, Song, Lee, and Park}]{Kim2003}
Kim JY, Song JY, Lee EJ, et~al (2003) Rheological properties and
  microstructures of carbopol gel network system. Colloid Polymer Sci
  281:614--623

\bibitem[{Kwak et~al(2015)Kwak, Ahn, and Song}]{Kwak2015}
Kwak MS, Ahn HJ, Song KW (2015) Rheological investigation of body cream and
  body lotion in actual application conditions. Korea Aust Rheol J 27:241--251

\bibitem[{Lapasin et~al(1995)Lapasin, {De Lorenzi}, Pricl, and
  Torriano}]{Lapasin1995}
Lapasin R, {De Lorenzi} L, Pricl S, et~al (1995) Flow properties of
  hydroxypropyl guar gum and its long-chain hydrophobic derivatives. Carbohydr
  Polym 28:195--202

\bibitem[{Lefran{\c{c}}ois et~al(2015)Lefran{\c{c}}ois, Ibarboure, Payr{\'e},
  Gontier, Le~Meins, and Schatz}]{Schatz2015}
Lefran{\c{c}}ois P, Ibarboure E, Payr{\'e} B, et~al (2015) Insights into
  carbopol gel formulations: Microscopy analysis of the microstructure and the
  influence of polyol additives. J Appl Polymer Sci 132:42761

\bibitem[{Li et~al(2018)Li, Tan, and Li}]{Li2018}
Li H, Tan C, Li L (2018) Review of 3d printable hydrogels and constructs. Mater
  Des 159:20--38

\bibitem[{Lubrizol(2006)}]{Ultrez}
Lubrizol (2006) Carbopol ultrez 20 polymer.
  \url{https://www.lubrizol.com/Personal-Care/Products/Product-Finder/Products-Data/Carbopol-Ultrez-20-polymer},
  last accessed 17/09/2021

\bibitem[{Morrison(2001)}]{morrison2001}
Morrison FA (2001) Understanding Rheology. Raymond F. Boyer Library Collection,
  Oxford University Press

\bibitem[{Mukherji et~al(2014)Mukherji, Marques, and Kremer}]{Kremer2014}
Mukherji D, Marques CM, Kremer K (2014) Polymer collapse in miscible good
  solvents is a generic phenomenon driven by preferential adsorption. Nature
  Comm 5:4882

\bibitem[{Nazareth et~al(2020)Nazareth, Karapetsas, Sefiane, Matar, and
  Valluri}]{Prash2020}
Nazareth RK, Karapetsas G, Sefiane K, et~al (2020) Stability of slowly
  evaporating thin liquid films of binary mixtures. Phys Rev Fluids 5:104007

\bibitem[{Niedzwiedz et~al(2009)Niedzwiedz, Arnolds, Willenbacher, and
  Brummer}]{Brummer2009}
Niedzwiedz K, Arnolds O, Willenbacher N, et~al (2009) Capillary breakup
  extensional rheometry of yield stress fluids. Appl~Rheol 19:41969

\bibitem[{Nishiyama and Satoh(2000{\natexlab{a}})}]{Nishiyama2000b}
Nishiyama Y, Satoh M (2000{\natexlab{a}}) Solvent- and counterion-specific
  swelling behavior of poly(acrylic acid) gels. J Polymer Sci B: Polymer Phys
  38:2791--2800

\bibitem[{Nishiyama and Satoh(2000{\natexlab{b}})}]{Nishiyama2000a}
Nishiyama Y, Satoh M (2000{\natexlab{b}}) Swelling behavior of poly(acrylic
  acid) gels in aqueous ethanol – effects of counterion species and ionic
  strength. Macromol Rapid Comm 21:174--177

\bibitem[{Oppong et~al(2006)Oppong, Rubatat, Frisken, Bailey, and
  de~Bruyn}]{Frisken2006}
Oppong FK, Rubatat L, Frisken BJ, et~al (2006) Microrheology and structure of a
  yield-stress polymer gel. Phys Rev E 73:041405

\bibitem[{Persson et~al(2013)Persson, Kovalev, and Gorb}]{Persson2013}
Persson BNJ, Kovalev A, Gorb SN (2013) Contact mechanics and friction on dry
  and wet human skin. Tribol Lett 50:17--30

\bibitem[{Poon et~al(2020)Poon, Brown, Direito, Hodgson, Le~Nagard, Lips,
  MacPhee, Marenduzzo, Royer, Silva, Thijssen, and Titmuss}]{Poon2020}
Poon WCK, Brown AT, Direito SOL, et~al (2020) Soft matter science and the
  covid-19 pandemic. Soft Matter 16:8310--8324

\bibitem[{Sappidi and Natarajan(2016)}]{Sappidi2016}
Sappidi P, Natarajan U (2016) Polyelectrolyte conformational transition in
  aqueous solvent mixture influenced by hydrophobic interactions and hydrogen
  bonding effects: {PAA}–water–ethanol. J Mol Graph Model 64:60 -- 74

\bibitem[{Tamura et~al(2013)Tamura, Ohsaki, and Nabata}]{Tamura2013}
Tamura E, Ohsaki K, Nabata Y (2013) Normal force as moisture feel for lipsticks
  - the control of spinnability. J Soc Rheol Jpn 41:223--227. In Japanese with
  English abstract

\bibitem[{Villa and Russo(2021)}]{Villa2021}
Villa C, Russo E (2021) Hydrogels in hand sanitizers. Materials 14:1577

\bibitem[{Warren(2017)}]{Warren2017}
Warren PB (2017) Sliding friction in the hydrodynamic lubrication regime for a
  power-law fluid. J Phys: Condens Matter 29:064005

\bibitem[{{World Health Organization}(2010)}]{WHO}
{World Health Organization} (2010) Guide to local production: {WHO}-recommended
  handrub formulations.
  \url{https://www.who.int/gpsc/5may/Guide_to_Local_Production.pdf}, last
  accessed 27/08/2020

\end{thebibliography}

\end{document}